\documentclass[aps,prd,preprint,nofootinbib,amsmath,amssymb,superscriptaddress]{revtex4-1}
\pdfoutput=1
\usepackage{latexsym}
\usepackage{color}
\usepackage{url}
\usepackage{hyperref}
\usepackage[utf8x]{inputenc} 
\usepackage[usenames,dvipsnames,svgnames,table]{xcolor}
\usepackage{graphicx}
\usepackage{epsfig}  
\usepackage{epsf}   
\usepackage{bm}
\usepackage{subfig}
\usepackage[toc,page]{appendix}
\usepackage{textcomp}
\usepackage{float}
\usepackage{hypcap}
\usepackage[]{hyperref}
\usepackage{booktabs}
\usepackage{multirow}
\usepackage{diagbox}
\usepackage{alphalph}
\usepackage{textcomp}
\usepackage{epstopdf}
\DeclareUnicodeCharacter{2212}{-}
\usepackage{lineno}
\usepackage{soul} 
\usepackage[normalem]{ulem}

\newcommand{\beq}{\begin{equation}}
\newcommand{\eeq}{\end{equation}}
\newcommand{\beqa}{\begin{eqnarray}}
\newcommand{\eeqa}{\end{eqnarray}}
\definecolor{orcidlogocol}{HTML}{A6CE39}

\begin{document}

\title{Study of light sterile neutrino at the long-baseline experiment options at KM3NeT} 

\author{Dinesh Kumar Singha}
\email{dinesh.sin.187@gmail.com}
\affiliation{School of Physics, University of Hyderabad, Hyderabad - 500046, India}

\author{Monojit Ghosh}
\email{mghosh@irb.hr}
\affiliation{School of Physics, University of Hyderabad, Hyderabad - 500046, India}
\affiliation{Center of Excellence for Advanced Materials and Sensing Devices, Ruder Bo\v{s}kovi\'c Institute, 10000 Zagreb, Croatia}

\author{Rudra Majhi}
\email{rudra.majhi95@gmail.com}
\affiliation{School of Physics, University of Hyderabad, Hyderabad - 500046, India}

\author{Rukmani Mohanta}
\email{rmsp@uohyd.ac.in}
\affiliation{School of Physics, University of Hyderabad, Hyderabad - 500046, India}

\begin{abstract}
 
 In this paper, we study the capability of different long-baseline experiment options at the KM3NeT facility i.e., P2O, Upgraded P2O and P2SO to probe the light sterile neutrino and compare their sensitivities with DUNE. The P2O option will have neutrinos from a 90 KW beam at Protvino to be detected at the ORCA detector, the Upgraded P2O will have neutrinos from the upgraded 450 KW beam to be detected at the ORCA detector and the option P2SO will have neutrinos from a 450 KW beam to be detected at the upgraded Super-ORCA detector. All these options will have a baseline around 2595 km. Our results show that the experiments at the KM3NeT (DUNE) would be more sensitive if the value of $\Delta m^2_{41}$ is around 10 (1) eV$^2$. Our results also show that the role of near detector is very important for the study of sterile neutrinos and addition of near detector improves the sensitivity as compared to only far detector for 3+1 scenario. Among the three options at KM3NeT, the sensitivity of P2O and upgraded P2O is limited and sensitivity of P2SO is either comparable or better than DUNE. 
 
\end{abstract}


\maketitle

\section{Introduction}
\label{intro}

The phenomenon of Neutrino oscillation  has been well established by variety of experiments like solar, atmospheric, reactor, accelerator etc. As a consequence of neutrino oscillation, neutrinos have to be massive possessing non-zero but small masses. Now we have entered  the era of precision measurement of neutrino oscillation parameters along with the determination of the remaining unknowns in standard three flavour scenario~\cite{Esteban:2020cvm}.  Neutrino experiments thus, provide an window to explore physics beyond the standard three flavour scenario. There are several such formulations exist in the literature, whose signatures can be probed through neutrino oscillation experiments. The possibility of having  more than three  neutrinos in nature has been discussed since a very long time among the physics community. If there exists a fourth neutrino, then it has to be sterile which can mix with the other three active neutrinos. Several experimental anomalies also hint towards the existence of an extra sterile neutrino~\cite{Acero:2022wqg}. Some crucial anomalies being the backbone of sterile neutrinos are LSND~\cite{LSND:2001aii} anomaly where a significant excess events have been seen over the known backgrounds, MiniBooNE~\cite{MiniBooNE:2020pnu} low energy excess while charged pion decay during flights, reactor \cite{Giunti:2021kab} and Gallium~\cite{Acero:2007su} anomalies associated with an overall normalization discrepancy of electron antineutrinos. The recent results of MicroBooNE~\cite{MicroBooNE:2021tya} have sparked the discussion of the light sterile neutrinos once again. According to MicroBooNE, there is no excess of $\nu_e$ events coming from the $\nu_\mu$ beam. However, Ref.~\cite{Denton:2021czb} studies the same set of MicroBooNE data and shows that an analysis with the electron disappearance channel is consistent with oscillations governed by sterile neutrinos. In reply to that, MicroBooNE performed a joined fit taking the $\nu_e$ appearance and $\nu_e$ disappearance channel and show that three neutrino fit still gives a better fit than four neutrino fit at $1 \sigma$ \cite{MicroBooNE:2022wdf}. In addition, a combined fit of MiniBooNE and MicroBooNE shows that the 3+1 model is still allowed at a significant confidence level~\cite{MiniBooNE:2022emn}. Apart from these accelerator  results, the gallium-based experiment BEST~\cite{Barinov:2021asz} and reactor based experiment Neutrino-4~\cite{Serebrov:2020kmd} recently reported a positive signal for a light sterile neutrino. On the other hand, the atmospheric data from the IceCube experiment is consistent with the no sterile neutrino hypothesis~\cite{IceCube:2020phf}. From the above discussions, we understand that the situation with a light sterile neutrino is still very intriguing and we need future data for arriving at a conclusive decision.

In this paper, we would like to study the sensitivity to the light sterile neutrinos at the long-baseline options of the KM3NeT facility~\cite{KM3Net:2016zxf}. The original aim of this facility is to detect the astrophysical neutrinos but recently there are discussions to send a neutrino beam from the accelerator in Protvino, Russia and detect these neutrinos at the detector at the KM3NeT facility~\cite{Akindinov:2019flp}. In this case, the distance from the source to the detector will be around 2595 km and energy of the neutrinos will be a few GeV. At this moment,  three options of this experiment are under consideration:  (i) neutrinos arising from a 90 KW proton beam from the accelerator at Protvino to be detected at the ORCA detector at the KM3NeT facility. We call this configuration as P2O, (ii) neutrinos generating from an upgraded 450 KW beam to be detected at ORCA detector. We call this configuration as Upgraded P2O and (iii) neutrino arising from an updated 450 KW beam to be detected at the updated Super-ORCA detector. We call this configuration as P2SO. The upgraded 450 KW beam as compared to 90 KW beam will produce a more intense beam of the neutrinos whereas an updated Super-ORCA detector over the ORCA detector will provide better efficiency to accept signal events and reject background events. In this work, we will consider all the three configurations as mentioned earlier to estimate the sensitivity of these experiments to a light sterile neutrino and compare their sensitivity with another upcoming long-baseline experiment DUNE~\cite{DUNE:2020ypp}. In our work, we consider two scenarios: (i) assuming light sterile neutrino does not exist in nature, we will estimate the upper bound on the sterile mixing parameters and (ii) assuming that a light sterile neutrino exists in nature, we will estimate how the sensitivity of these experiments to measure the current unknowns in standard three flavour scenario is getting affected. We consider both far and near detectors in our analysis. To the best of our knowledge, this is the first work estimating the sensitivity to sterile neutrinos at the long-baseline options at the KM3NeT facility. For previous studies regarding the study of light sterile neutrinos in the context of long-baseline experiments, we refer to Refs.~\cite{Berryman:2015nua,Gandhi:2015xza,Agarwalla:2016xxa,Agarwalla:2016xlg,Penedo:2022etl,Coloma:2017ptb,Choubey:2017cba,Choubey:2017ppj,Haba:2018klh,Ghosh:2017atj,KumarAgarwalla:2019blx,Majhi:2019hdj,Reyimuaji:2019wbn,Ghosh:2021rtn,Ghosh:2019zvl,Denton:2022pxt,Dutta:2016glq,Choubey:2018kqq,Donini:2007yf,Capozzi:2016vac}. 

The paper is organized as follows. In the next section we discuss the 3+1 scenario i.e., how the oscillation in the standard three flavour is altered by the existence of one light sterile neutrino. In Sec.~\ref{spec} we will discuss the configuration of the different experimental setups that we consider in our study. We will also briefly outline the statistical procedure which we adopt to estimate the sensitivity. In Sec.~\ref{prob}, we discuss the sensitivities of different experiments to the sterile neutrinos at the probability level. In Sec.~\ref{bound}, we estimate the bound on the sterile mixing parameters assuming there is no sterile neutrino in nature. After that, in Secs.~\ref{hier}, \ref{oct}, \ref{cpv} and \ref{cpp}, we will study the sensitivity of these experiments to measure the current unknowns in the standard three flavour scenario assuming there exists a light sterile neutrino in nature. Finally, in Sec.~\ref{sum} we summarize our results and then conclude. 

\section{Formalism}

In presence of a light sterile neutrino, the  PMNS mixing matrix $U$ which relates the  neutrino flavour states to the mass  states, is written in the following way:
\begin{eqnarray}
&&~~~~~~~~~~~U =  U_{34}(\theta_{34},\delta_{34})
U_{24}(\theta_{24},\delta_{24})
U_{14}(\theta_{14},0)
U^{3\nu}\,,  \label{prm} \\
&&{\rm with}~~~~~~U^{3\nu}
=
U_{23}(\theta_{23},0)
U_{13}(\theta_{13},\delta_{\rm CP})
U_{12}(\theta_{12},0)\,,
\end{eqnarray}
where $U_{ij}(\theta_{ij},\delta_{ij}$) denotes a rotation in the $(i,j)$-plane with mixing angle $\theta_{ij}$ and phase $\delta_{ij}$\footnote{ Note that throughout our text we have used the notation $\delta_{\rm CP}$ for $\delta_{13}$.}.
Neutrino oscillation in the standard three flavour is governed by three mixing angles: $\theta_{13}$, $\theta_{12}$ and $\theta_{23}$, two mass squared differences: $\Delta m^2_{21} = m_2^2 - m_1^2$ and $\Delta m^2_{31} = m_3^2 - m_1^2$ and one phase $\delta_{\rm CP}$. Among these parameters, the current unknowns are: (i) hierarchy of the neutrino masses, (ii) octant of $\theta_{23}$ and (iii) the CP violating phase $\delta_{\rm CP}$. In the presence of a sterile neutrino, this mixing scheme is extended by three new mixing angles: $\theta_{14}$, $\theta_{24}$ and $\theta_{34}$, two additional phases: $\delta_{24}$ and $\delta_{34}$ and one more mass squared difference: $\Delta m^2_{41}$. 

  At this point, let us briefly discuss our choice of the parametrization of $U$ is 3+1 scenario. Note that in Eq.~\ref{prm}, the phases can be associated with any of the three mixing angles. For study in short-baseline experiments where the oscillations are mainly governed by $\Delta m^2_{41}$, it would be sufficient to associate all the three phases in $U^{3\nu}$ and therefore, in the probability expressions there will be no phases \cite{Donini:2007yf}. As we are interested in the long-baseline regime, the phases become important. In case of oscillations governed by $\Delta m^2_{31}$ only, the ideal way is to put the two phase with $\theta_{12}$ and, $\theta_{13}$ and to put the other phase with any of the sterile angles \cite{Donini:2007yf}.  For the $\nu_\mu \rightarrow \nu_e$ oscillations at the long-baseline, the oscillations are governed by both $\Delta m^2_{21}$ and $\Delta m^2_{31}$. Therefore, in this case, it is important to associate the sterile phases with the sterile angles \cite{Capozzi:2016vac}. However, the mixing angle $\theta_{34}$ appears in the oscillation probabilities only when one considers strong matter effect and neutral current events in the calculation, which we will see in the next paragraph. Therefore, keeping one phase with $\theta_{34}$, one can easily neglect the effect of $\theta_{34}$ and $\delta_{34}$ in the analysis, if one does not consider matter effect and neutral current events.

The appearance channel probability (P($\nu_{\mu} \rightarrow \nu_{e}$)) is an important channel to verify the active neutrino oscillations to sterile neutrino  at the near detector. As our beam consists of muon type of neutrinos, we do not expect any significant signal of electron type of neutrinos above the known background as the neutrinos at the near detector will be unoscillated due to insufficient distance for oscillation to occur. If significant excess of electron type of neutrinos are detected at the near detector, it may indicate the presence of sterile neutrino, that  oscillates to active electron type of neutrinos. The appearance channel probability of neutrinos in effective two flavor approximation with an extra light sterile neutrino at the near detector can be found in Refs. \cite{Ghosh:2019zvl, Boser:2019rta, Klop:2014ima,Kopp:2013vaa}:
 \begin{align}
 \label{ap_prob}
  P^{\rm ND}_{\mu e} = \sin^{2} 2\theta_{14} \sin^{2} \theta_{24} \sin^{2} \left( \frac{\Delta m^{2}_{41} L}{4 E_{\nu}}  \right),
 \end{align} 
 where $\theta_{1 4}$ and $\theta_{2 4}$ are sterile mixing angles, $\Delta m^{2}_{4 1}$ is the sterile mass squared difference, $E_{\nu}$ is neutrino energy and $L$ is the baseline length. As we can see the probability term is dependent on the sterile oscillation parameters only, hence the near detector facilitates the best platform to study the sterile neutrinos. 
 
 The approximate appearance channel probability at the far detector which was calculated for the first time in Ref.~\cite{Klop:2014ima}, is given by 
 
 \begin{eqnarray}
 \label{ap_probfd}
  P^{\rm FD}_{\mu e} &=&  4 s^{2}_{13} s^{2}_{23} \sin^{2} {\Delta_{31}} \\ \nonumber
  &+& 8 s_{12} c_{12} s_{13} s_{23} c_{23} \sin \Delta_{21} \sin \Delta_{31} \cos \left( \Delta_{31} + \delta_{\rm CP} \right)\\ \nonumber
  &+& 4 s_{13} s_{14} s_{24} s_{23} \sin \Delta_{31} \sin \left( \Delta_{31} + \delta_{\rm CP} + \delta_{24} \right),
 \end{eqnarray}
 where $\Delta_{ij} = \frac{\Delta m^{2}_{ij} L}{4E}$. To arrive at this expression one needs to use the long baseline approximations ($\Delta_{21} \rightarrow 0$ and  $\Delta_{41} \rightarrow \infty$). It should be noted that the probability expression in equation \ref{ap_probfd} does not depend on $\Delta m^{2}_{41}$. The reason being,  for the $L/E$ values relevant for far detectors, the oscillations are averaged out for any particular value of $\Delta m^{2}_{41}$. Therefore, this probability expression is independent of $\Delta m^{2}_{41}$. Again this expression is independent of $\theta_{34}$ and $\delta_{34}$. But this won't be the case when we consider the interaction with matter as both $\theta_{34}$ and $\delta_{34}$ will be modified due to matter effect, affecting the oscillation probability. However, the parameters, $\theta_{34}$ and $\delta_{34}$ become more effective if the neutral current events are considered. This behavior has been first pointed out in the analytical treatment presented in Ref.~\cite{Klop:2014ima} and successively verified in the numerical simulations performed in Ref.~\cite{Gandhi:2017vzo}.

\section{Experimental setup and simulation details}
\label{spec}

We have used GLoBES~\cite{Huber:2004ka,Huber:2007ji} software package to simulate DUNE and three configurations at KM3NeT. We have further used additional plugins of GLoBES to implement the sterile neutrino in our study~\cite{Kopp:2006wp}.

\begin{figure}[hbt!]
\begin{center}
\includegraphics[width=0.49\textwidth]{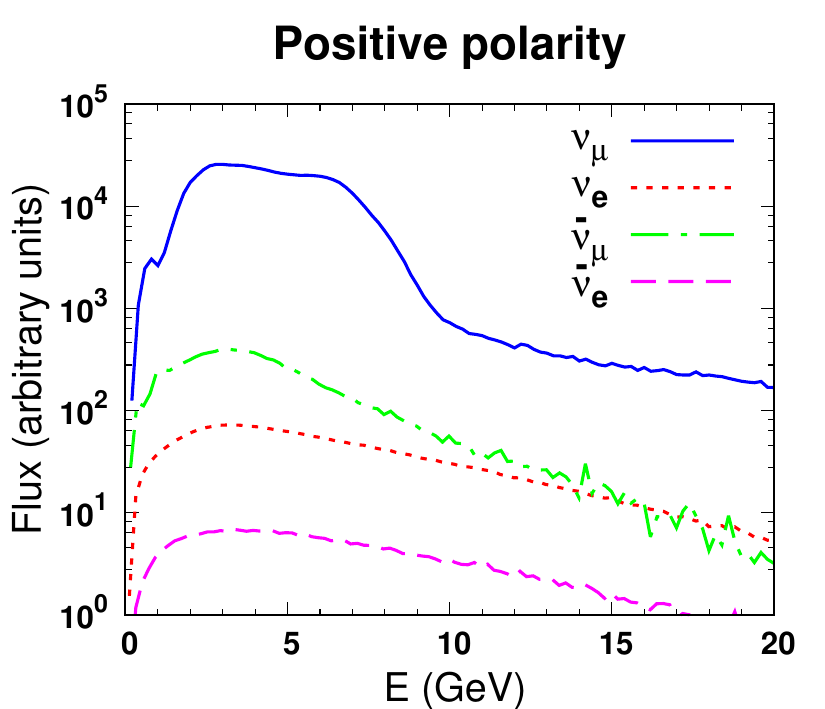}
\includegraphics[width=0.49\textwidth]{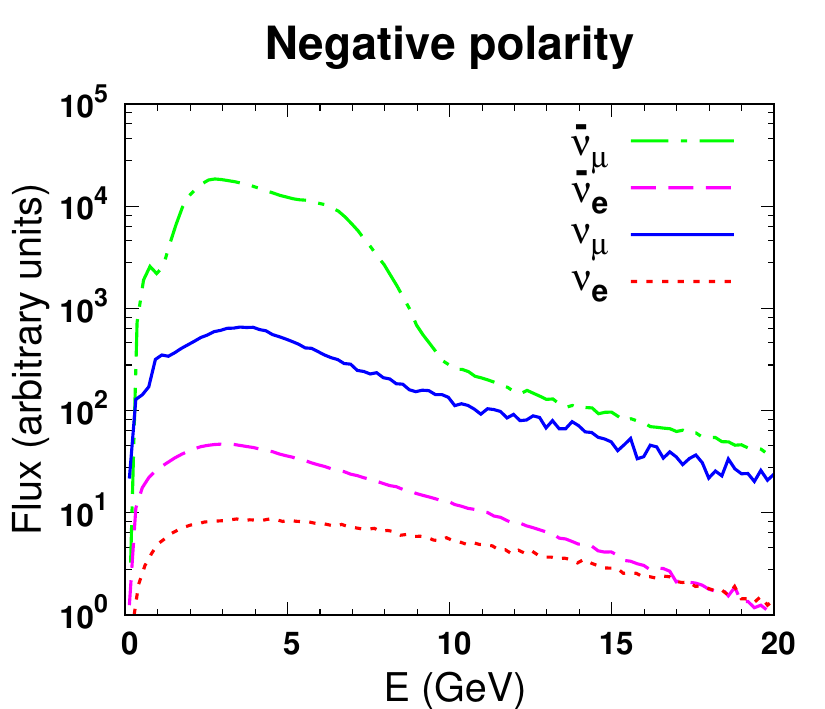}
\end{center}
\caption{Fluxes (in arbitrary units) as a function of neutrino energy for P2O experiment.}
\label{fig_flux}
\end{figure}

For simulating the long-baseline experiments at KM3NeT, we have taken the three configurations from Refs.~\cite{Akindinov:2019flp,Hofestadt:2019whx}. The Protvino accelerator contains a 1.5 km circumference U-70 synchrotron which produces 90 KW beam corresponding to 0.8 $\times$ $10^{20}$ protons on target (POT) per year for P2O and 450 KW beam corresponding to 4 $\times$ $10^{20}$ POT per year for Upgraded P2O and P2SO configurations. For P2O and Upgraded P2O experiments, the neutrinos will be detected at ORCA (Oscillation Research with Cosmics in the Abyss) far detector (FD) located in the Mediterranean Sea about 40 km off the coast of Toulon, France. It is located at a depth between 2450 m (the seabed depth) and 2250 m. The ORCA detector will contain 8 Mt sea water in total out of which 4 Mt would be its fiducial volume. This detector will be about 2595 km away from the neutrino source. We used the fluxes as given in Ref. \cite{Akindinov:2019flp} and shown in Fig.~\ref{fig_flux}. In this figure, we have shown the components of $\nu_\mu$/$\bar{\nu}_\mu$, $\nu_e$/$\bar{\nu}_e$ fluxes in both positive and negative polarities. From these panels, we understand that the flux corresponding to signal (i.e., $\nu_\mu$ in the positive polarity and $\bar{\nu}_\mu$ in the negative polarity) is higher as compared to flux corresponding to background. However, in Ref.~\cite{Singha:2021jkn}, we showed that for P2O, the major background for the $\nu_e$ signal events arise from $\nu_\mu$ misidentified as $\nu_e$, neutral current (NC) events identified as $\nu_e$ and $\nu_\tau$ identified as $\nu_e$. This is because, the ORCA detector is mainly designed to detect atmospheric neutrinos and ultra high energy neutrinos from the extra-galactic sources. The electron events in the ORCA detector produce a shower. At 5 GeV, the efficiency of $\nu_e$ events classified as shower is 90\%. But there are also 78\% of the $\nu_\tau$ events, 85\% of the NC events and 50\% of the $\nu_\mu$ events  classified as shower
which act as a background for the $\nu_e$ selection \cite{KM3Net:2016zxf}. We have included these backgrounds in our analysis. We employed pre-smearing energy dependant efficiencies to match the energy spectrum shown in Fig. 7 of Ref~\cite{Akindinov:2019flp}. These efficiencies account for the energy-dependent effective mass of the ORCA detector, as shown in Fig. 90 of Ref. \cite{KM3Net:2016zxf}. The energy resolution and particle identification factor were obtained from Figs. 68 and 99 of Ref. \cite{KM3Net:2016zxf}. We have checked that our physics sensitivity matches with Figs. 8 and 10 of Ref.~\cite{Akindinov:2019flp} corresponding to P2O and Upgraded P2O, respectively. 

The Super-ORCA detector for the P2SO experiment will be 10 times denser than the ORCA detector. In comparison to ORCA, the Super-ORCA detector would have a lower energy threshold for neutrino detection, improved neutrino flavour identification capacity, and higher energy resolution. Using pre-smearing energy dependant efficiencies, we matched the energy spectrum of P2SO as shown in Fig. 5 of Ref. \cite{Hofestadt:2019whx} and the particle identification factor was taken from Fig. 1. Our results reproduce the physics sensitivity of P2SO as given in Figs. 13, 14 and 15 of Ref.~\cite{Akindinov:2019flp}. We have considered a total run-time of six years, divided into three years in neutrino mode and three years in antineutrino mode for all three configurations of the long-baseline options at KM3NeT. The energy window for calculation of events is 2 GeV to 12 GeV for P2O/Upgraded P2O and 0.2 GeV to 10 GeV in P2SO.  

We utilised the official GLoBES files corresponding the technical design report~\cite{DUNE:2021cuw} for DUNE. A 40 kt liquid argon time-projection chamber detector with a beam power of 1.2 MW and a running time of 7 years is positioned 1300 km away from the source, producing $1.1 \times 10^{21}$ POT per year. The run-time has been divided into 3.5 years in neutrino mode and 3.5 years in antineutrino mode. The neutrino source will be located in Fermilab in the United States, and the detector will be located in South Dakota. 

Now let us discuss about the configuration of the near detectors (ND). For DUNE, we have considered an identical detector as that of far detector having volume of 0.147 kt at a distance of 574 m from the source. For KM3NeT, we have used a DUNE like near detector placed at a distance of 320 m from the source. Here we have taken the detector volume in such a way that the number of signal events at the near detector for DUNE and the number of signal events at the near detector for P2O match with each other for $\Delta m^2_{41} = 1$ eV$^2$. 
 
This way we ensure that the sensitivities coming from the near detector of DUNE and near detector of P2O are comparable. 

\begin{table} 
\centering
\begin{tabular}{|c|c|} \hline
Parameters            & True values $\pm$ $1\sigma$       \\ \hline
$\sin^2 \theta_{12}$  & $0.304^{+ 0.013}_{- 0.012}$      \\ 
$\sin^2 \theta_{13}$ & $0.0222^{+ 0.00068}_{- 0.00062}$                 \\ 
$\sin^2 \theta_{23} $ & $0.573^{+ 0.018}_{- 0.023}$                 \\ 
$\delta_{\rm CP}[^\circ] $       & $194^{+ 52}_{- 25}$         \\ 
$\Delta m^2_{21}$ [10$^{-5}$ eV$^2$]    & $7.42^{+ 0.21}_{- 0.20}$  \\ 
$\Delta m^2_{31}$ [10$^{-3}$ eV$^2$]   & $2.515^{+ 0.028}_{- 0.028}$    \\ \hline
$\sin^2\theta_{14}$ & 0.0076 \\
$\sin^2\theta_{24}$ & 0.0076 \\
$\sin^2\theta_{34}$ & 0  \\
$\delta_{24} [^\circ]$ &0 \\
$\delta_{34} [^\circ]$ &0  \\
$\Delta m^2_{41}$ [eV$^2$]   & 0.1, 1, 10 \\ 

 \hline
\end{tabular}
\caption{The values of oscillation parameters that we considered in our analysis. Standard oscillation parameters considered from~\cite{Esteban:2020cvm} with their corresponding 1$\sigma$ errors.}
\label{table_param}
\end{table}  

For the estimation of the sensitivity, we use the Poisson log-likelihood and assume that it is $\chi^2$-distributed:
\begin{equation}
 \chi^2_{{\rm stat}} = 2 \sum_{i=1}^n \bigg[ N^{{\rm test}}_i - N^{{\rm true}}_i - N^{{\rm true}}_i \log\bigg(\frac{N^{{\rm test}}_i}{N^{{\rm true}}_i}\bigg) \bigg]\,,
\end{equation}
where $N^{{\rm test}}$ is the number of events in the test spectrum, $N^{{\rm true}}$ is the number of events in the true spectrum and $i$ is the number of energy bins. The best-fit values of the oscillation parameters and their $1 \sigma$ errors are taken from NuFIT~\cite{Esteban:2020cvm} and are listed in Table~\ref{table_param}. In our analysis, we have marginalized over all the relevant parameters using the built-in minimizer in GLoBES taking a prior corresponding to their $1 \sigma$ uncertainty. For $\delta_{CP}$ and the sterile oscillation parameters, we have not considered any priors and they are allowed to vary freely. The systematic is incorporated by the method of pull \cite{Fogli:2002pt,Huber:2002mx}. For systematic errors, we have taken an overall normalization and shape errors corresponding to signal and background. We list the values of the systematic errors for KM3NeT and DUNE in Table \ref{table_sys}. It should be noted that the DUNE GLoBES file contains no shape errors.  We show all our results for the normal hierarchy of the neutrino masses.

\begin{table} 
\centering
\begin{tabular}{|c|c|c|c|c|} \hline
Systematics            & P2O         & Up P2O    & P2SO          & DUNE  \\ \hline
Sg-norm $\nu_{e}$  & 7$\%$   &  7$\%$  & 5$\%$   & 2$\%$      \\ 
Sg-norm $\nu_{\mu}$ & 5$\%$ &  5$\%$   & 5$\%$            & 5$\%$ \\ 
Bg-norm & 12$\%$    &   12$\%$   & 12$\%$     & 5$\%$ to 20$\%$\\ 
Sg-shape & 11$\%$    & 11$\%$       & 11$\%$     & NA\\ 
Bg-shape &  4\% to 11\%  & 4$\%$ to 11 \%      & 4\% to 11$\%$       & NA\\ 
\hline
\end{tabular}
\caption{The values of systematic errors that we considered in our analysis. ``norm" stands for normalization error, ``Sg" stands for signal and ``Bg" stands for background.}
\label{table_sys}
\end{table}  


\section{Discussion at the probability level}
\label{prob}

 \begin{figure}[hbt!]
\begin{center}
\includegraphics[width=0.49\textwidth]{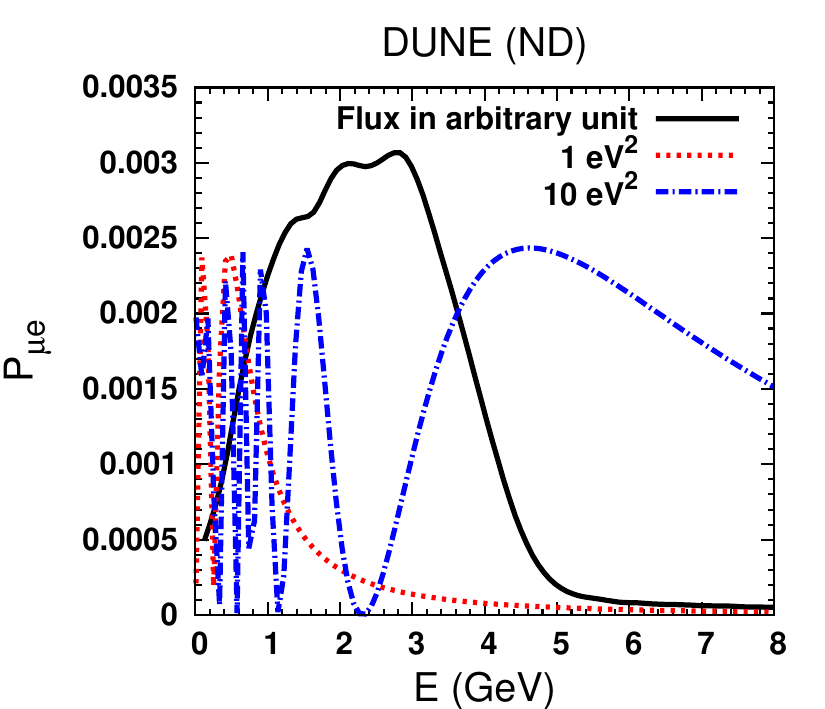}
\includegraphics[width=0.49\textwidth]{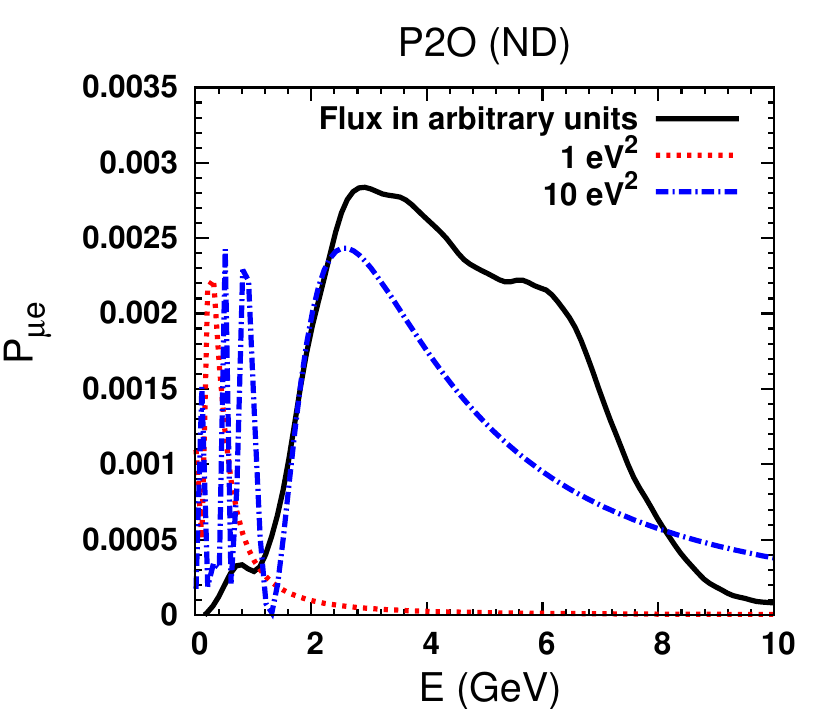} 
\end{center}
\caption{Electron neutrino appearance probability as a function of energy for DUNE and P2O near detectors for $\Delta m^{\rm 2}_{41}$ as 1 eV$^2$ and 10 eV$^2$. Also we have shown the corresponding fluxes of the experiments in probability plot.}
\label{fig_prob}
\end{figure}

In Fig. \ref{fig_prob}, we have shown the appearance channel probability as a function of neutrino energy for different values of $\Delta m^{2}_{4 1}$. The left plot is for DUNE near detector which is placed at 574 meters from the neutrino beam source and the right plot is for P2O near detector which is placed at 320 meters from the source. The red and blue  curves correspond to $\Delta m^{2}_{4 1} = $ 1 eV$^2$ and 10 eV$^2$, respectively. The black curves correspond to the neutrino fluxes in arbitrary units. Both the experiments are insensitive to sterile neutrino mass squared difference of $\Delta m^{2}_{4 1} = 0.1$ eV$^2$, which we have not shown here. Next we see that most part of the red curve lies inside the flux envelop for DUNE near detector but this is not the case for the P2O near detector. From this we can draw an important conclusion that the DUNE experiment will be more sensitive to $\Delta m^{2}_{4 1} = 1$ eV$^2$ sterile neutrinos. The first maxima for $\Delta m^{2}_{4 1} = 10$ eV$^2$ coincides with the flux maxima of P2O experiment whereas second maxima and some part of first maxima lies inside the flux envelop of DUNE experiment. Therefore, we can expect that P2O experiment will provide best sensitivity to $\Delta m^{2}_{4 1} = 10 $ eV$^2$ sterile neutrinos.

\section{Constraining sterile oscillation parameters}
\label{bound}

In this section, we will estimate the bound on the different sterile mixing parameters assuming there are no light sterile neutrinos in nature.

\begin{figure}[hbt!]
\begin{center}
\includegraphics[width=0.325\textwidth]{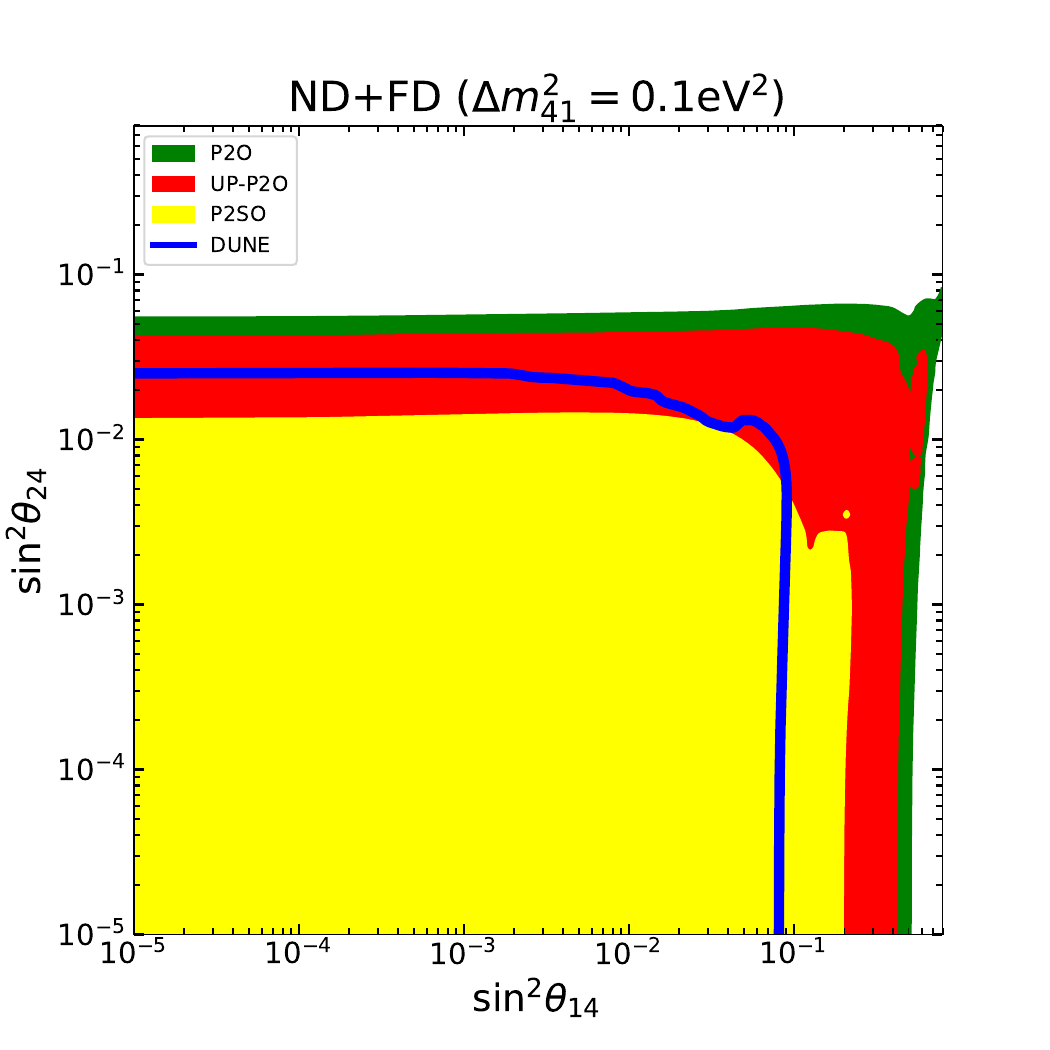}
\includegraphics[width=0.325\textwidth]{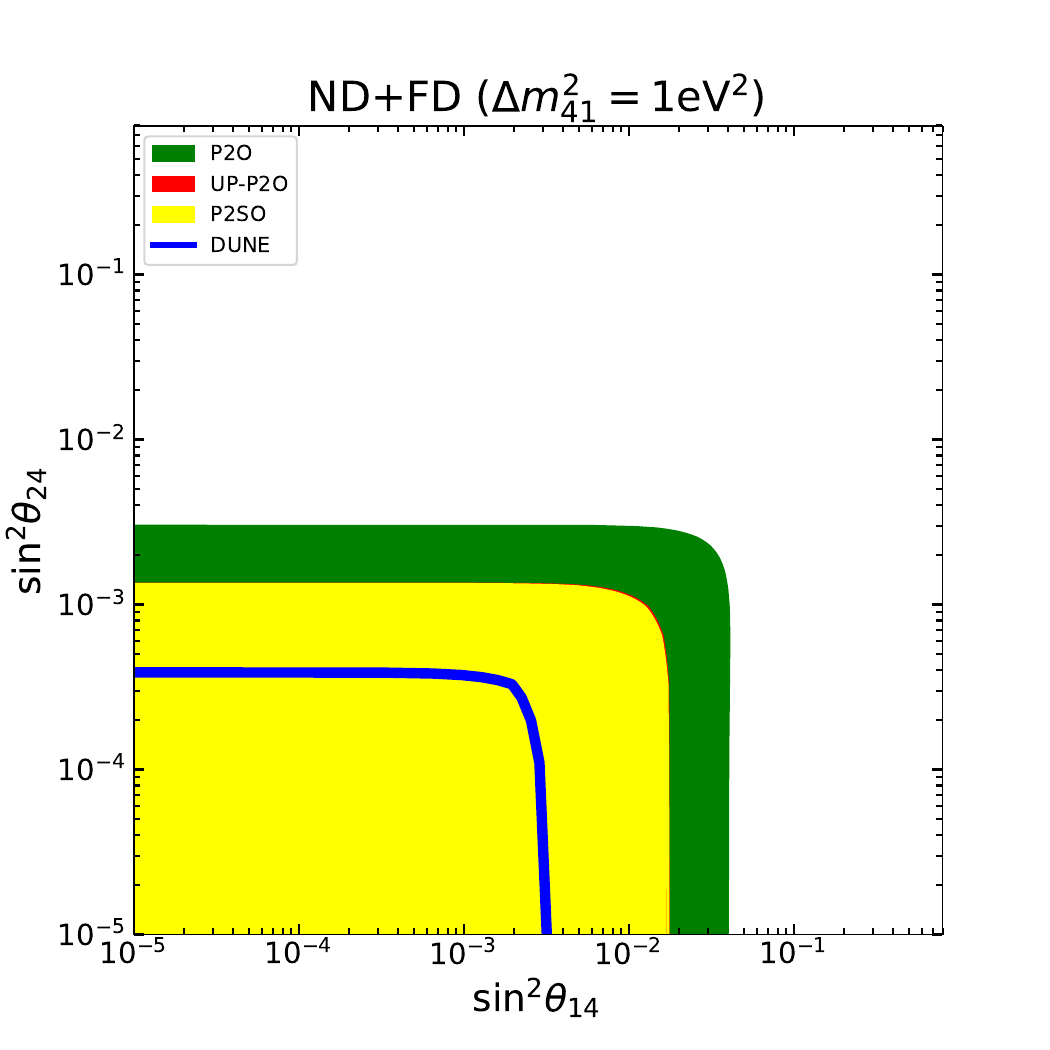} 
\includegraphics[width=0.325\textwidth]{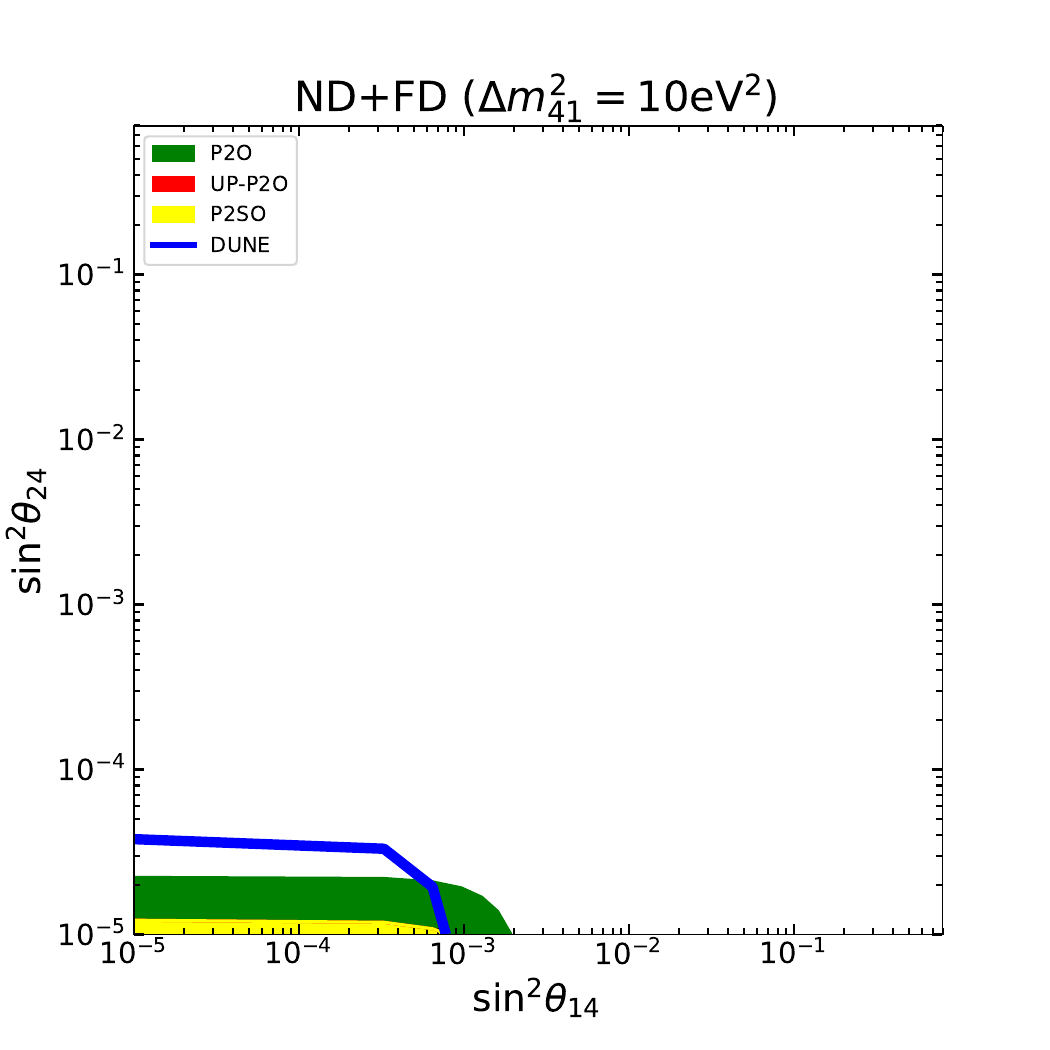}
\includegraphics[width=0.325\textwidth]{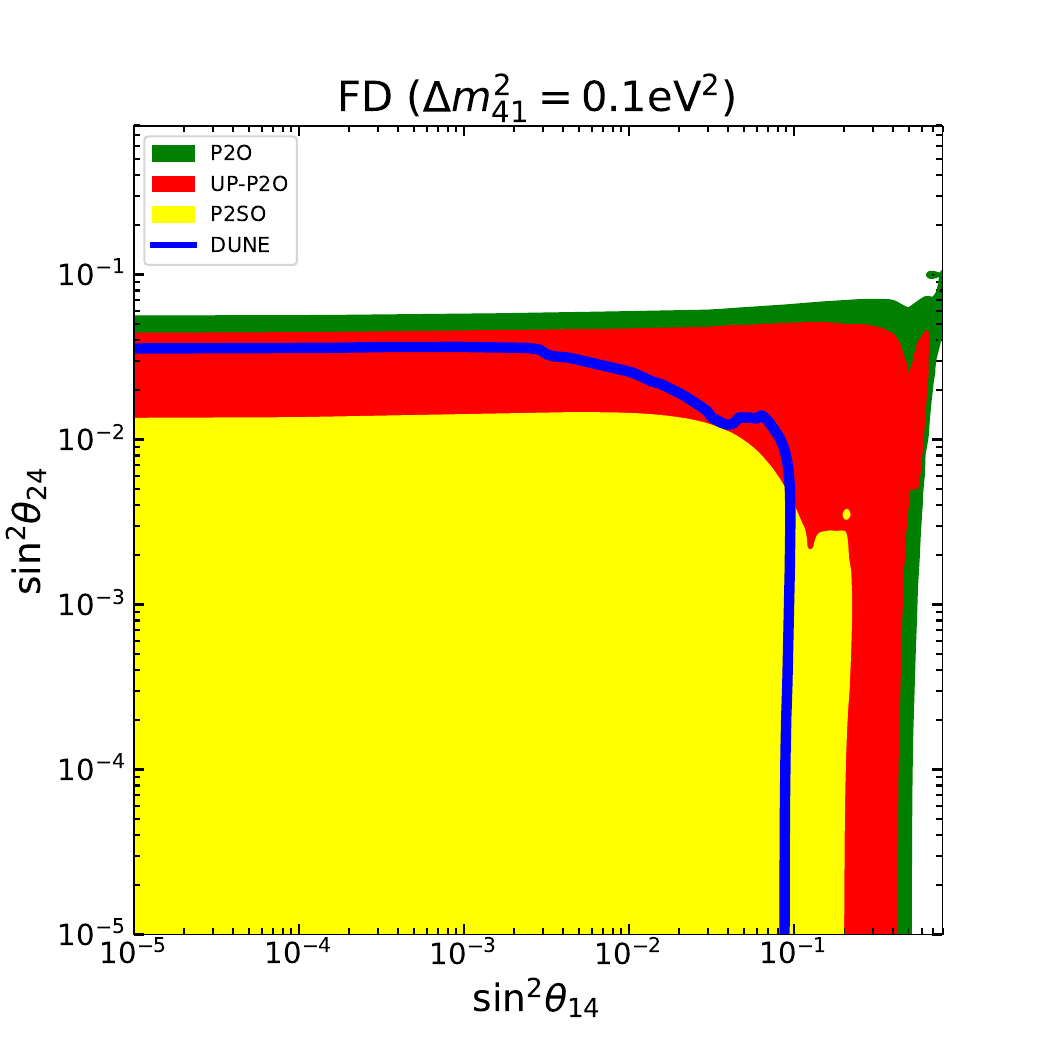}
\includegraphics[width=0.325\textwidth]{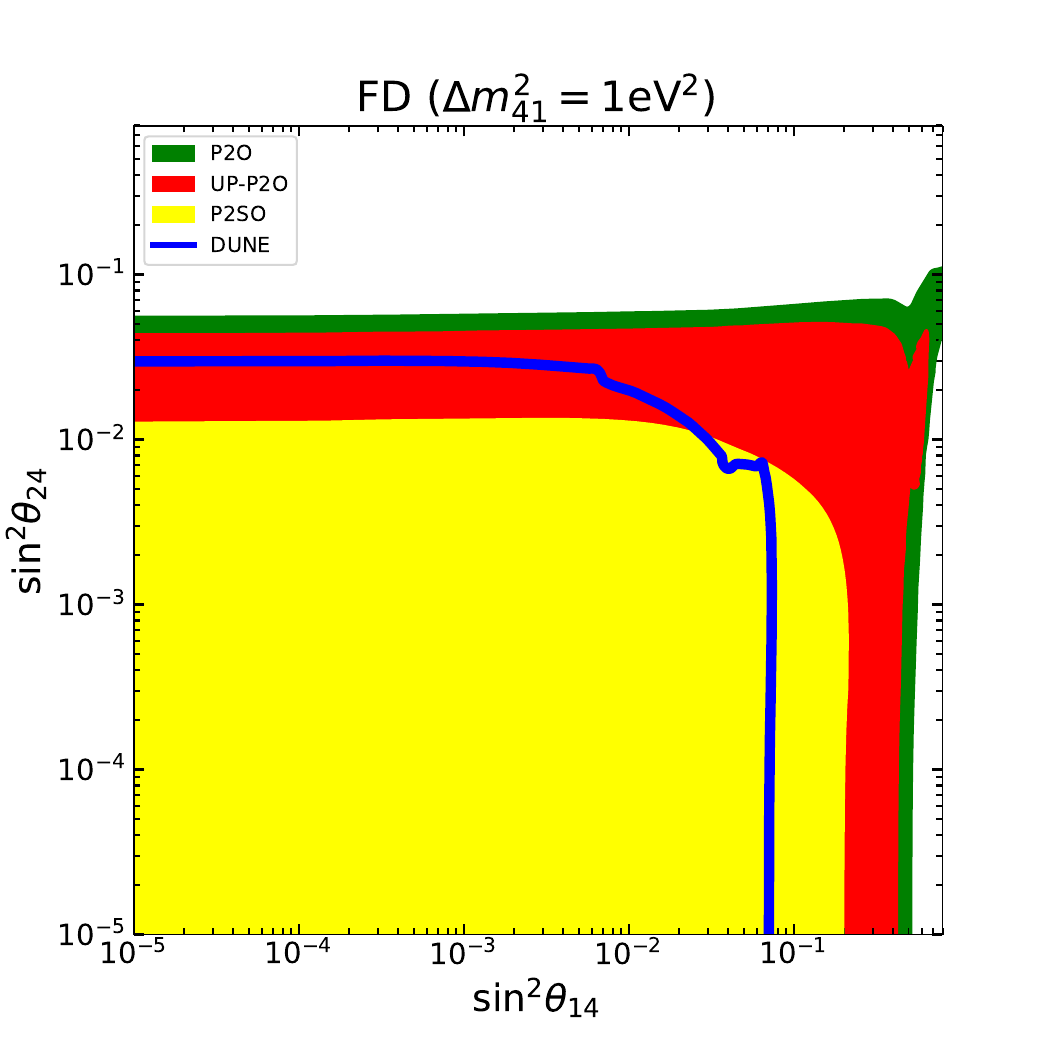} 
\includegraphics[width=0.325\textwidth]{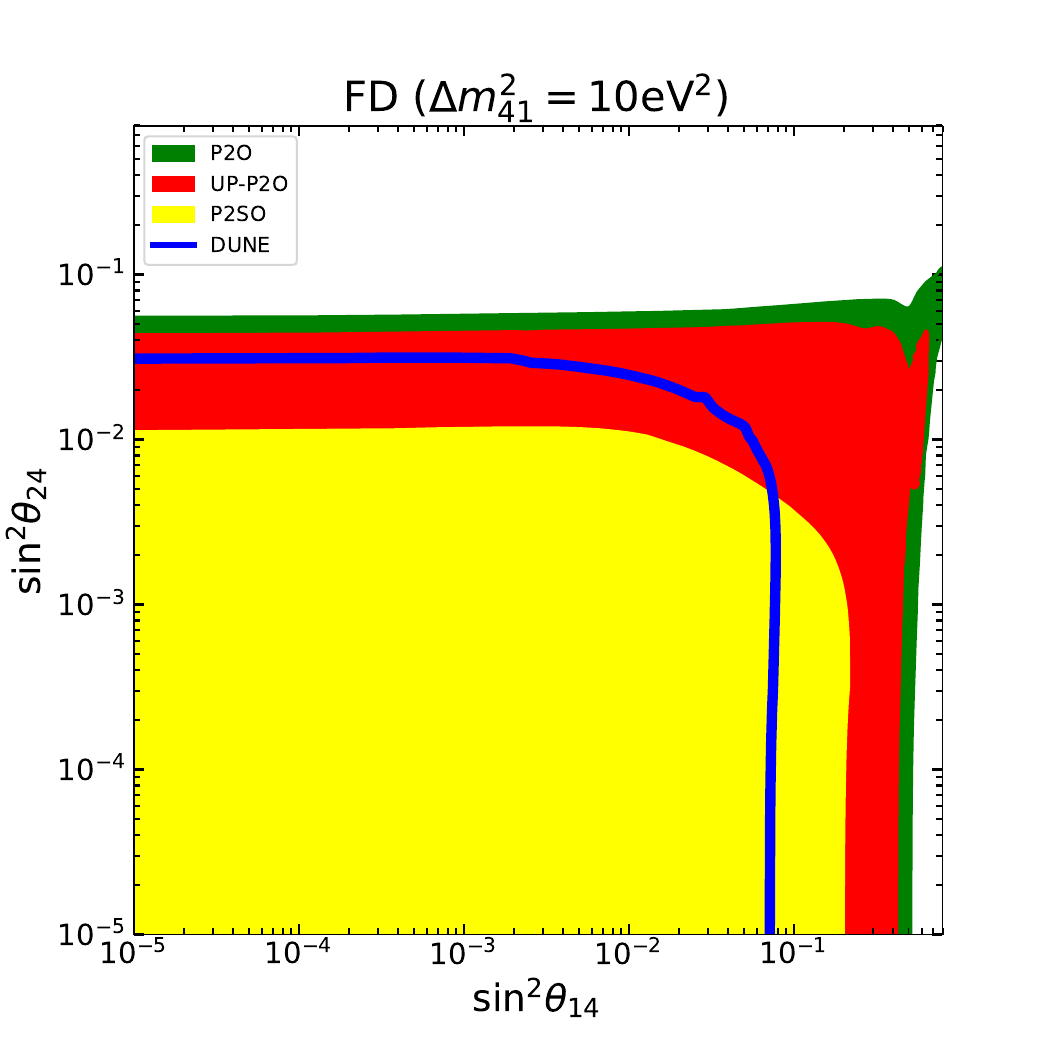}
\end{center}
\caption{Capability of DUNE, P2O, Upgraded P2O and P2SO to constrain the sterile neutrino parameters $\theta_{14}$ and $\theta_{24}$. In the upper (lower) panel we present the two-dimensional sensitivity curves in $\sin^2 \theta_{1 4}$ - $\sin^2 \theta_{2 4}$ plane for ND+FD (FD) at 95$\%$ C.L.}
\label{fig_bound}
\end{figure}

In Fig. \ref{fig_bound}, we have shown the allowed region for the sterile oscillation parameters $\sin^{2} \theta_{1 4}$ and $\sin^{2} \theta_{2 4}$ at 95\% confidence level. The upper (lower) panel corresponds to the case where ND+FD (FD) configuration was considered for constraining these parameters. The different colors correspond to various experiments. The blue curve represents the allowed region for DUNE experiment. While green, red, and yellow bands represent the regions for P2O, Upgraded P2O, and P2SO experiments, respectively. Left, middle and right columns in the figure represent three different values of the sterile mass squared differences $\Delta m^{2}_{4 1} = 0.1 $ eV$^2$, 1  eV$^2$ and 10 eV$^2$, respectively. If we compare the lower panel with the upper panel, we can identify that the FD+ND configurations are giving more stringent bounds on the sterile mixing angles compared to the case where only FD is considered except the case for $\Delta m^{2}_{4 1} = 0.1$ eV$^2$. The improvement in the ND+FD case is due to the fact that NDs are more sensitive to oscillations due to sterile neutrino ($\Delta m^2_{41} = 1$ eV$^2$ and 10 eV$^2$) compared to the only FD cases due to the averaging of the oscillations frequency due to rapid oscillations. As discussed earlier, for $\Delta m^2_{41} = 0.1$ eV$^2$ NDs of P2O and DUNE experiments are insensitive to sterile neutrinos and for this reason, adding a ND does not help in the improvement in the sensitivity. If we focus on the case for $\Delta m^{2}_{41} = 1 $ eV$^2$, we can clearly see that DUNE experiment gives the best bound for the ND+FD configuration and if we shift our focus to the case for $\Delta m^{2}_{41} = 10 $ eV$^2$, we can see that P2SO and Upgraded P2O configurations are giving best bound on the sterile parameters. This observation can be well understood from Fig. \ref{fig_prob}. We can see one more interesting feature in these two plots that the P2SO and Upgraded P2O bands are overlapping for the ND+FD case. This is due to the fact that the NDs for both P2SO and Upgraded P2O are the same in terms of both beam exposure and running period. The sensitivity of P2O is inferior than Upgraded P2O and P2SO but better than DUNE for $\Delta m^2_{41} = 10$ eV$^2$ and FD+ND. In the case of FD only, as we change the value of $\Delta m^{2}_{41}$ we do not see any significant change in the bounds for all the experiments. This is due to the fact that the the appearance channel probability, Eqn (\ref{ap_prob}) does not depend on $\Delta m^{2}_{4 1}$ when we consider FD only. Note that the current bounds on the sterile mixing angles are available from the ongoing experiments \cite{Dentler:2018sju}. But our results can't be compared with theirs as they are obtained using a different parametrization of the PMNS matrix.

\section{Hierarchy sensitivity}
\label{hier}

\begin{figure}[hbt!]
\begin{center}
\includegraphics[width=0.49\textwidth]{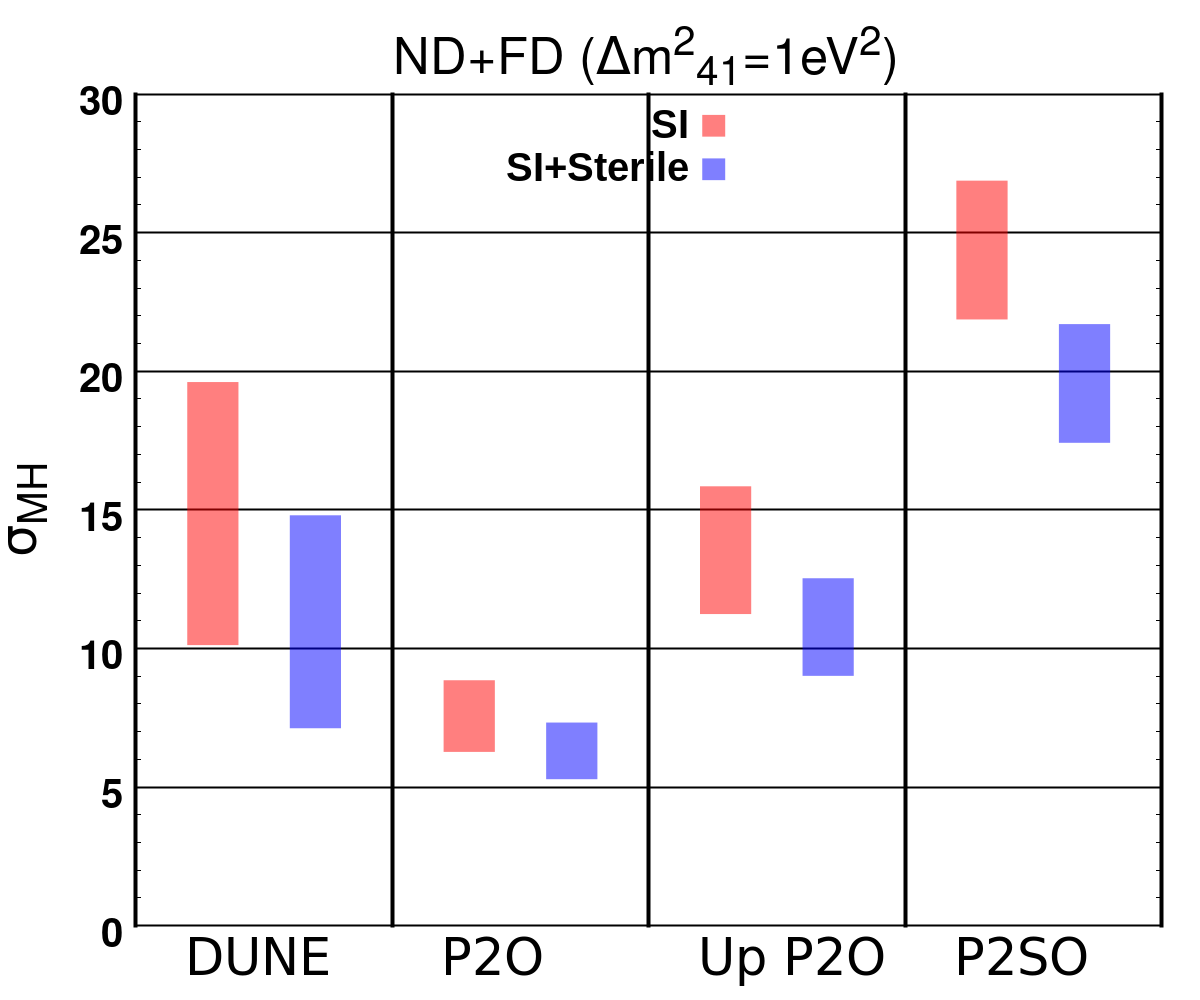}
\includegraphics[width=0.49\textwidth]{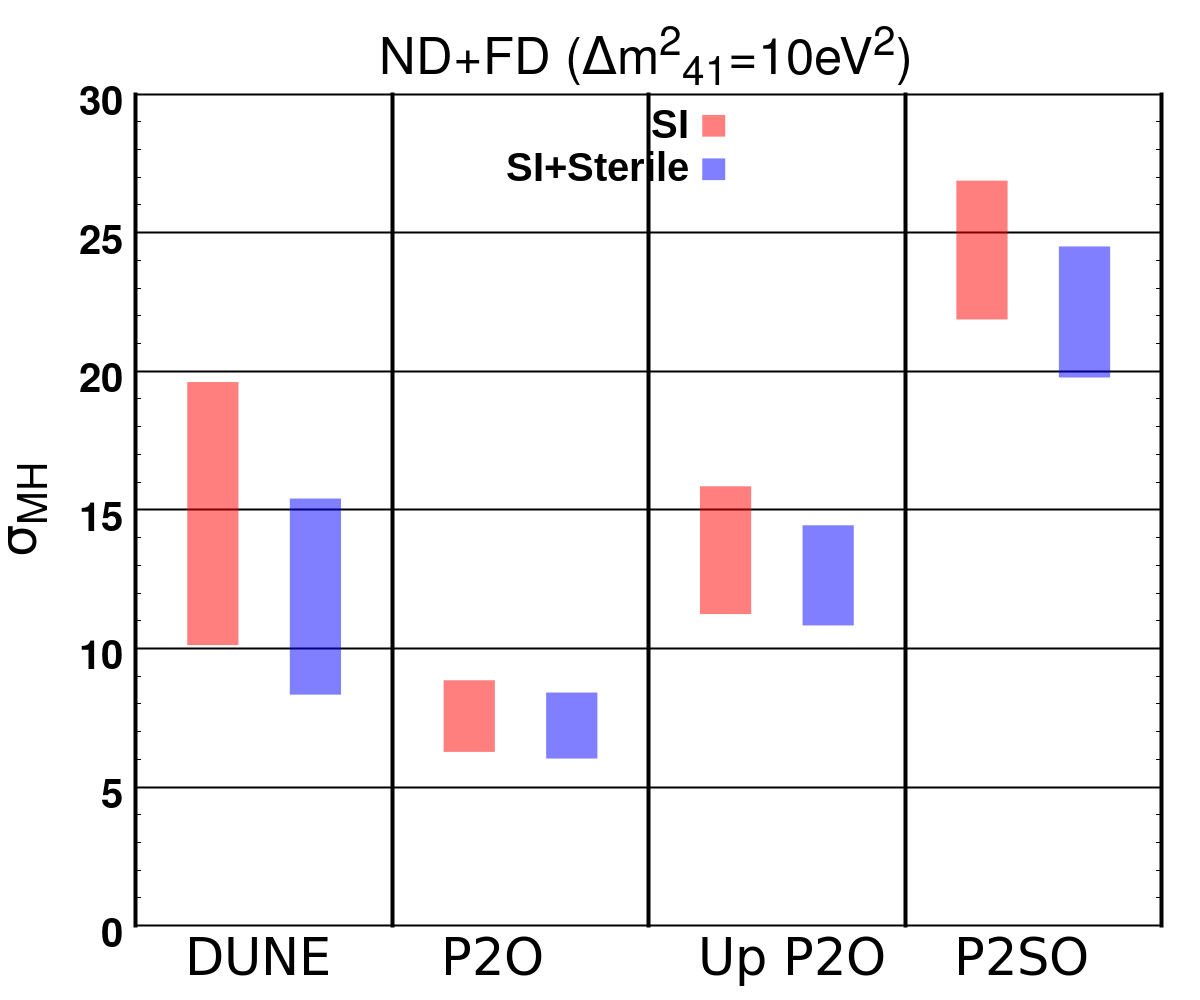} 
\includegraphics[width=0.49\textwidth]{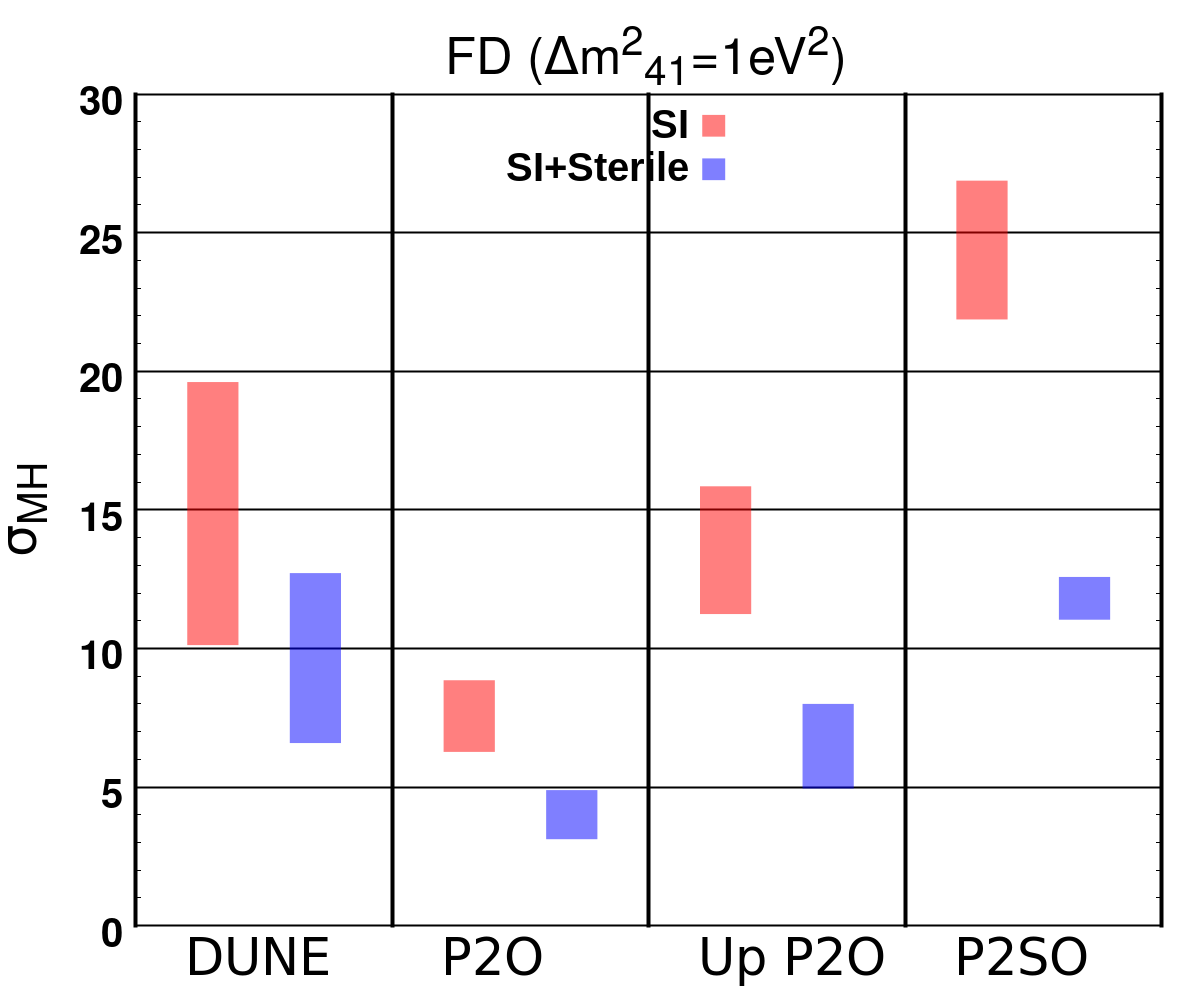}
\includegraphics[width=0.49\textwidth]{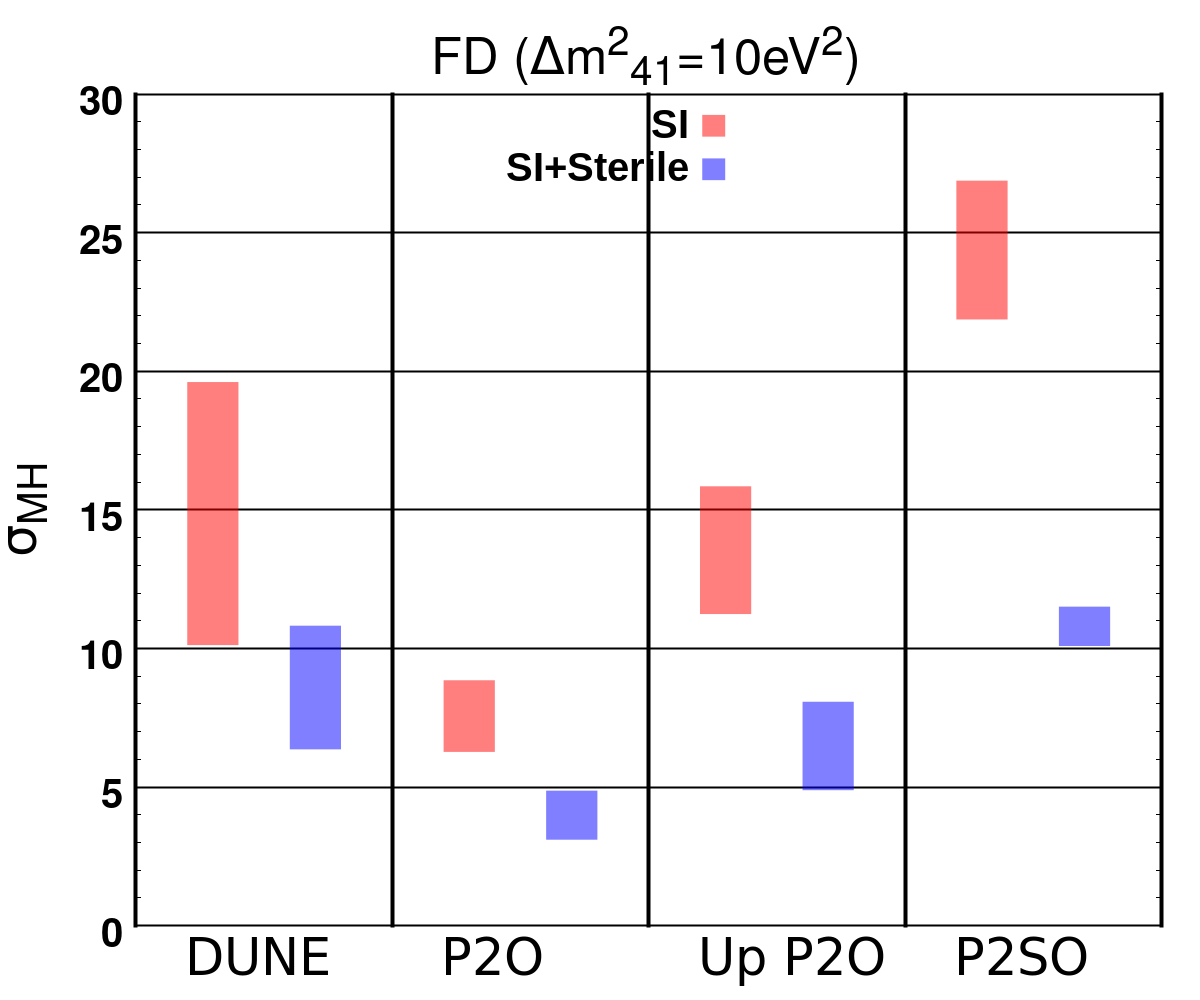}
\end{center}
\caption{Range of MH sensitivities of DUNE, P2O, Upgraded P2O (Up P2O) and P2SO for the standard interaction (SI) case as well as SI + sterile case for $\Delta m^{\rm 2}_{4 1} = 1$ eV$^2$ and 10  eV$^2$.}
\label{fig_hier}
\end{figure}

 The sensitivities to neutrino mass hierarchy for different long-baseline experiments in standard three flavor and in the presence of sterile neutrino have been shown in Fig \ref{fig_hier}. Each column represents the sensitivity from $\chi^2_{\rm min}$ to $\chi^2_{\rm max}$ arising due to the different values of true $\delta_{\rm CP}$. Hierarchy sensitivity is the capability of the experiment to exclude the wrong hierarchy. In order to calculate hierarchy sensitivity, we obtain $\chi^2$ assuming true hierarchy as NH and test hierarchy as IH. The upper panel represents the situation where both ND and FD were included for the study, whereas the lower panel corresponds to the FD case only. The red bars represent the hierarchy sensitivity in standard three flavor scenario and the blue bars represent the case with one extra sterile neutrino. In each row, the left panel corresponds to $\Delta m^2_{41} = 1$ eV$^2$ and the right panel corresponds to $\Delta m^2_{41} = 10$ eV$^2$. 

As we know the hierarchy sensitivity is directly proportional to the matter effect, i.e higher the baseline, higher the hierarchy sensitivity. Holding on to that logic, we would expect that P2O experiment (all three configurations) should provide more sensitivity to hierarchy. As we can see from the figure that in standard interaction case P2SO gives the best sensitivity to mass hierarchy, whereas DUNE and Upgraded P2O give the next best sensitivity. P2O is least sensitive to hierarchy due to its low background rejection capability and 5 time less beam power compared to Upgraded P2O and P2SO. We have shown this fact in our previous work \cite{Singha:2021jkn}. We can notice that the Upgraded P2O sensitivity to mass hierarchy is less than the P2SO sensitivity although their baseline and beam power are same. This is due to the fact that Upgraded P2O uses ORCA detector with limited background rejection capability compared to the improved Super-ORCA detector used by P2SO.

In the presence of a sterile neutrino, the hierarchy sensitivities of all the experiments decrease due to the degeneracies arising between standard oscillation parameters with the sterile mixing angles and sterile phases. As the FDs are not sensitive to the sterile neutrinos, the hierarchy sensitivities of all the experiments are low. There is no significant difference when we change the sterile mass squared difference $\Delta m^{2}_{4 1}$ from 1 eV$^2$ to 10 eV$^2$. Once we introduce the NDs, we find an improvement in their sensitivities. This is because, inclusion of ND provides a constraint on the sterile mixing parameters which are free in case of only FD. For P2O, Upgraded P2O and P2SO, the improvement in the sensitivity is more in the case of $\Delta m^2_{41} = 10$ eV$^2$ as compared to $\Delta m^2_{41} = 1$ eV$^2$. For DUNE, the improvement is same for both $\Delta m^2_{41} = 10$ eV$^2$ and 1 eV$^2$. The hierarchy sensitivity of Upgraded P2O in the presence of sterile neutrino is comparable to DUNE both for $\Delta m^{2}_{41} = 1 $ eV$^2$ and $\Delta m^{2}_{41} =$10 eV$^2$. Similar to the standard interaction case P2O gives the least sensitivity to hierarchy in the presence of sterile neutrino.  

\section{Octant sensitivity}
\label{oct}

\begin{figure}[hbt!]
\begin{center}
\includegraphics[width=0.49\textwidth]{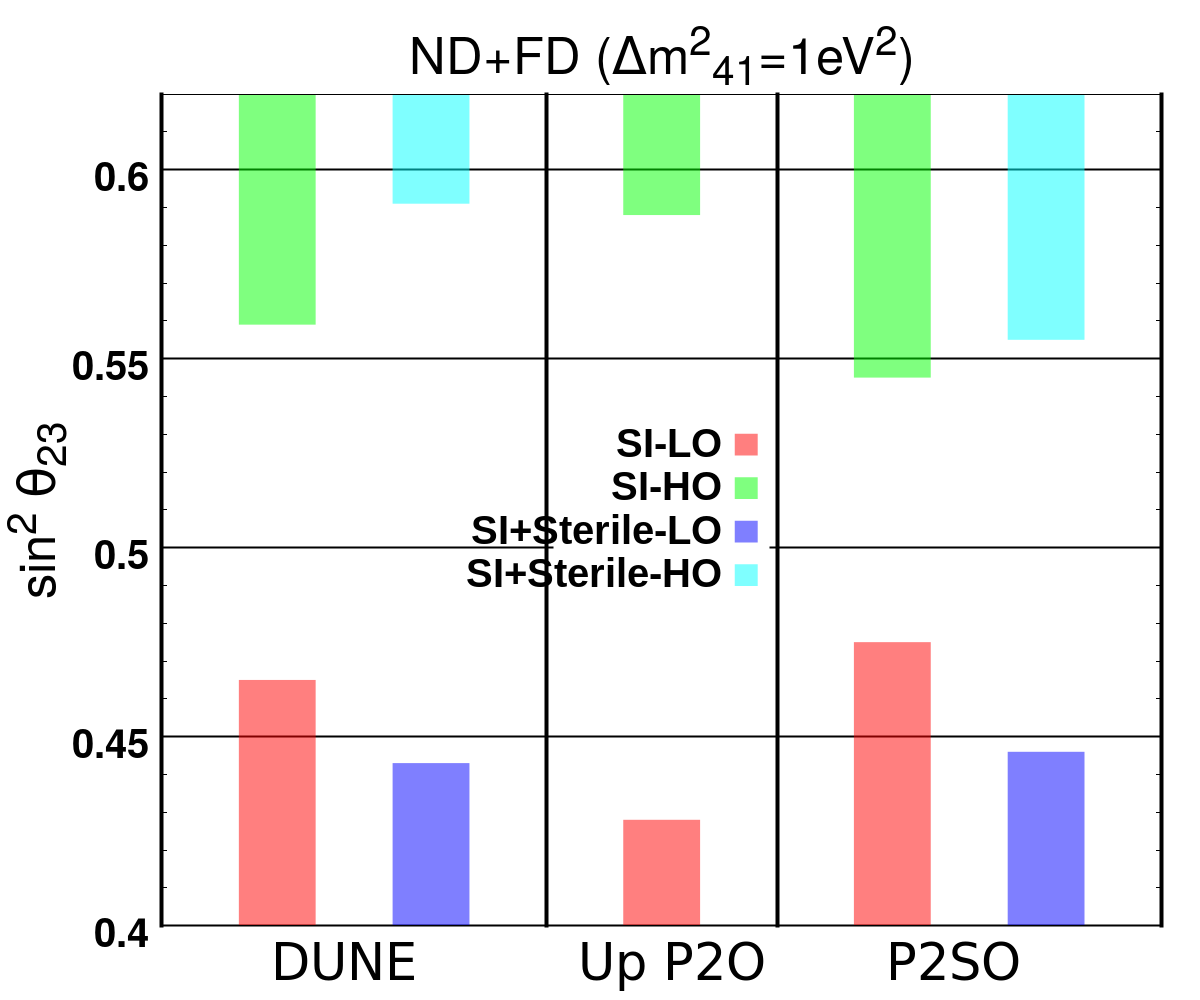}
\includegraphics[width=0.49\textwidth]{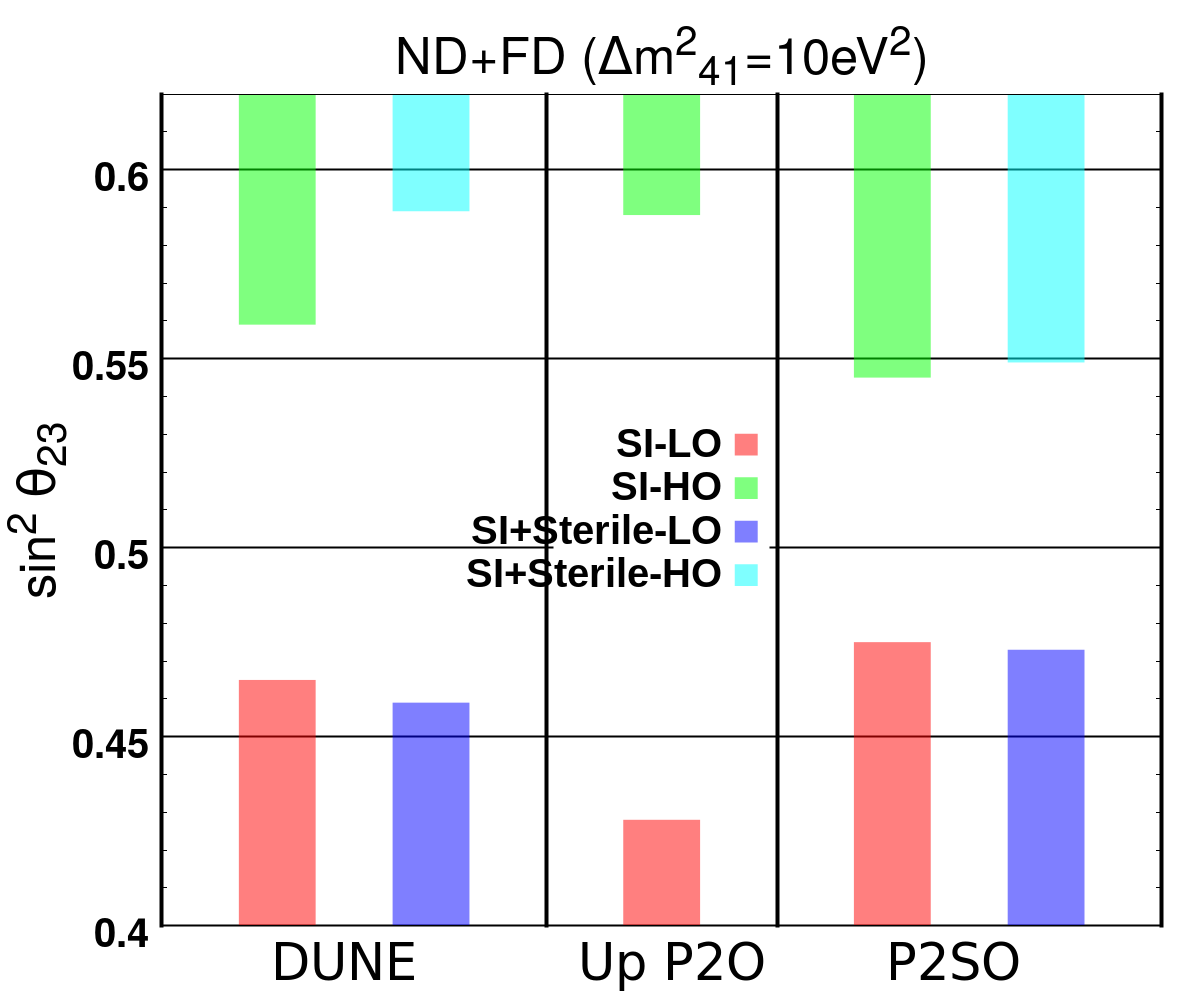} 
\includegraphics[width=0.49\textwidth]{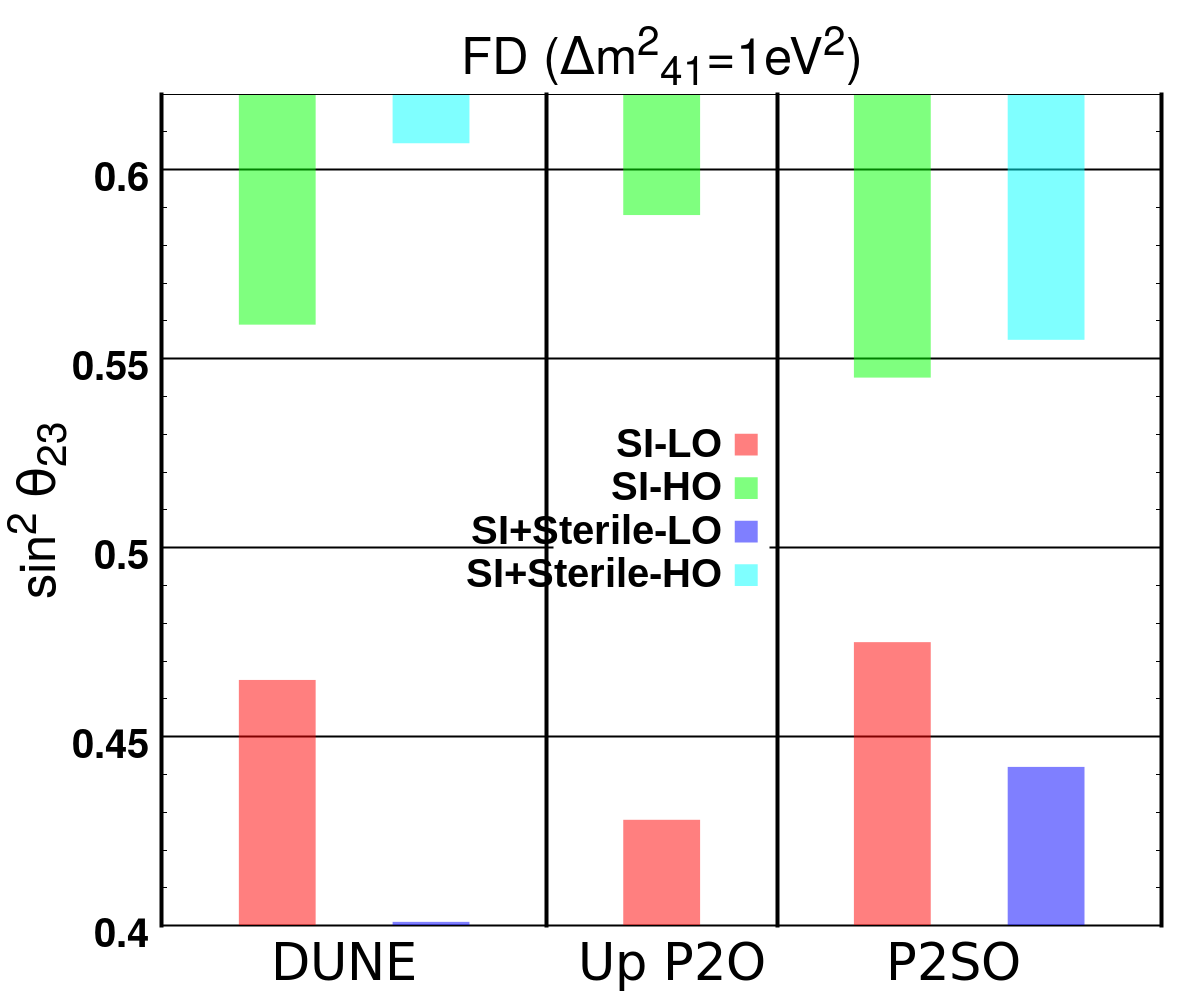}
\includegraphics[width=0.49\textwidth]{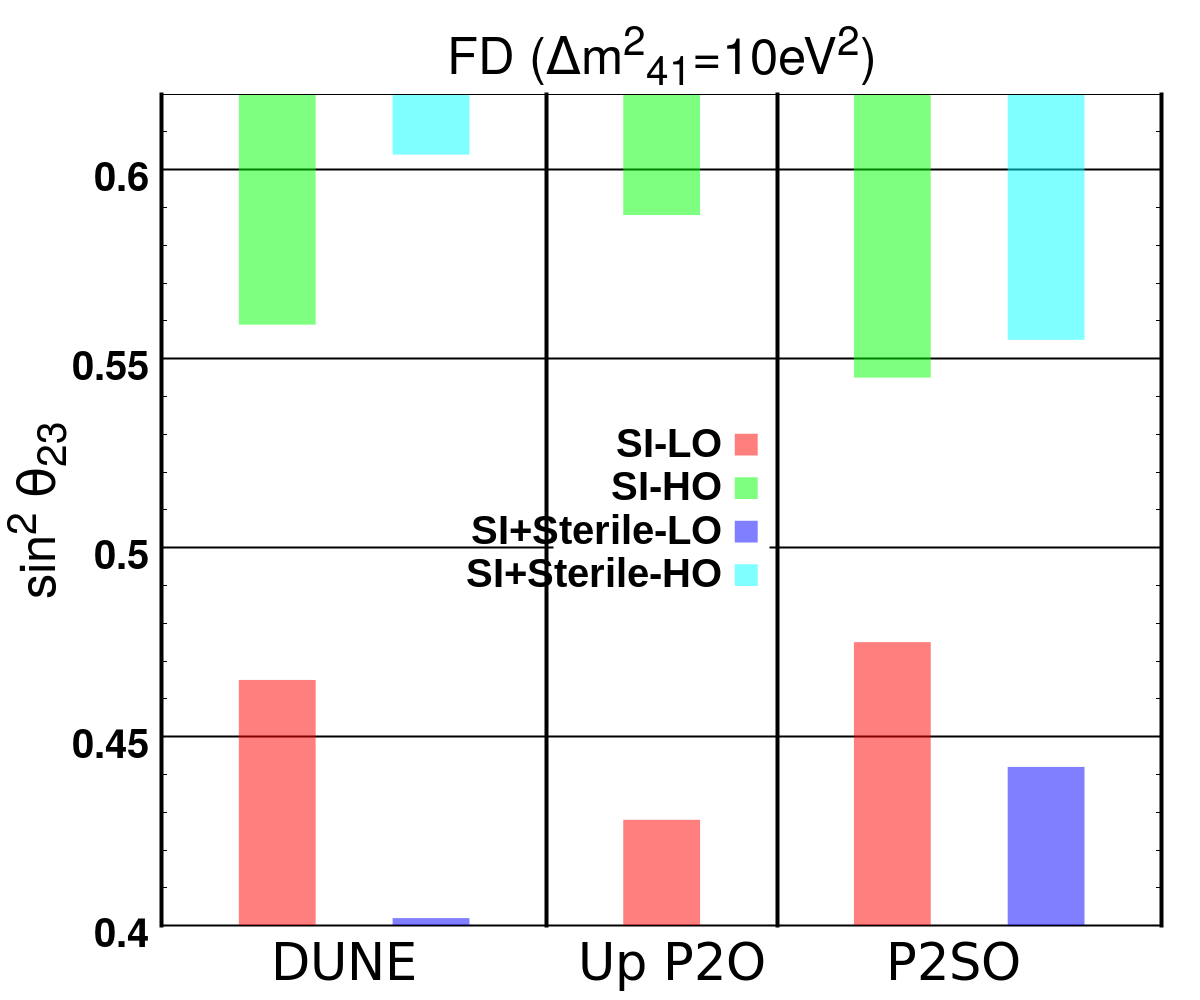}
\end{center}
\caption{Region of $\sin^2 \theta_{23}$ for which octant can be determined above $3\sigma$ C.L. for DUNE, Upgraded P2O (Up P2O) and P2SO experiments in standard interaction (SI) case and SI + sterile case for sterile neutrinos of masses $\Delta m^{\rm 2}_{4 1} =$  1 eV$^2$ and 10 eV$^2$.}
\label{fig_oct}

\end{figure}

  In Fig. \ref{fig_oct}, we have shown the region of $\sin^2 \theta_{23}$ in the range  [0.4:0.62], where the true octant can be determined above $3\sigma$ C.L. for DUNE, Upgraded P2O and P2SO experiments in the standard three flavour scenario and in presence of sterile neutrino. Octant sensitivity represents the capability of the experiment to exclude the degenerate solutions for $\theta_{23}$ if any exists. In order to calculate octant sensitivity, we obtain $\chi^2$ assuming the true $\theta_{23}$ in LO and varying the test $\theta_{23}$ in HO and vice versa. In each panel, red (blue) band represents the range of $\sin^2\theta_{23}$ in LO having octant sensitivity greater than $3\sigma$ in the standard (SI + sterile) scenario. Similarly green (cyan) band represents the range of $\sin^2\theta_{23}$ for HO in the standard (presence of sterile) neutrino scenario. 
Upper panel of the figure represents the sensitivities considering the combination of ND and FD for each long baseline experiment. While, plots in lower panel show the octant sensitivities with only FD. Each plot on the left panel displays octant sensitivities assuming the new mass squared difference $\Delta m^2_{41}$ as 1 eV$^2$, while the right panel plots are obtained for $\Delta m^2_{41}$ as 10 eV$^2$. 

 The sensitivities to octant degeneracy are relatively weak and below $3\sigma$ for almost whole range of $\theta_{23}$ for P2O  experiment, hence not shown in the figure. The Upgraded P2O experiment has some octant sensitivity above 3$\sigma$ only in standard interaction case, while no such sensitivity is obtained for sterile neutrino. DUNE and P2SO experiments are more sensitive to octant of $\theta_{23}$. One can easily observe deterioration in octant sensitivity with only FD for all the experiments after including the sterile neutrino to the standard three flavor scenario.  Another crucial observation is related to the contribution of ND in each experiment. As discussed earlier, ND is more important for analysis of sterile neutrino. Hence, inclusion of ND to FD in each experiment enhances the sensitivity significantly. But for P2SO experiment ( blue band) there is notable increment in  region for LO, while for HO region no such change in sensitivity is observed after including ND to FD as compared to only FD. In order to understand this anomalous result from P2SO, we have calculated the values of different oscillation parameters at the minimum $\chi^2_{\rm min}$ values corresponding to $\theta_{23}$ in LO and HO regions. Interestingly for $\theta_{23}$ in LO, we observed that the parameters $\theta_{34}$ and $\delta_{34}$ are more constrained in FD+ND compared to only FD. No such observations are found for HO. For further clarification we have removed the dependency on $\theta_{34}$ and $\delta_{34}$ by assuming their values to be zero and obtained no such anomalous octant sensitivities. This implies the fact that there exists some degeneracies involving the sterile parameters $\theta_{34}$ and $\delta_{34}$. As a result for LO, both $\theta_{34}$ and $\delta_{34}$ are unconstrained with only FD. ND helps to tightly constrain these parameters in LO regions. FD is quite effective to constrain $\theta_{34}$ and $\delta_{34}$ in HO region and hence we observed no such effect for ND in HO region.

\section{CP violation sensitivity}
\label{cpv}

\begin{figure}[hbt!]
\begin{center}
\includegraphics[width=0.49\textwidth]{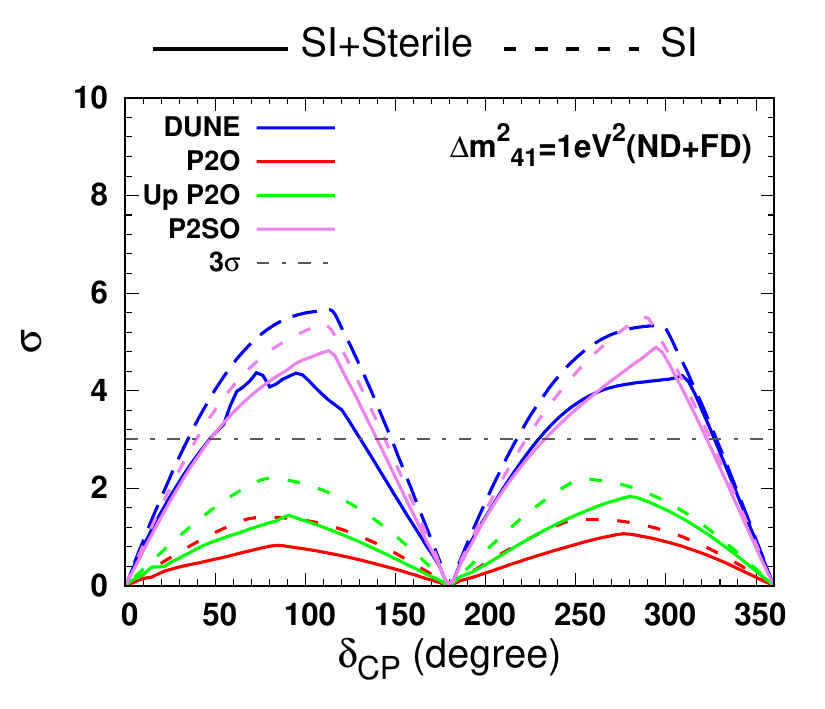}
\includegraphics[width=0.49\textwidth]{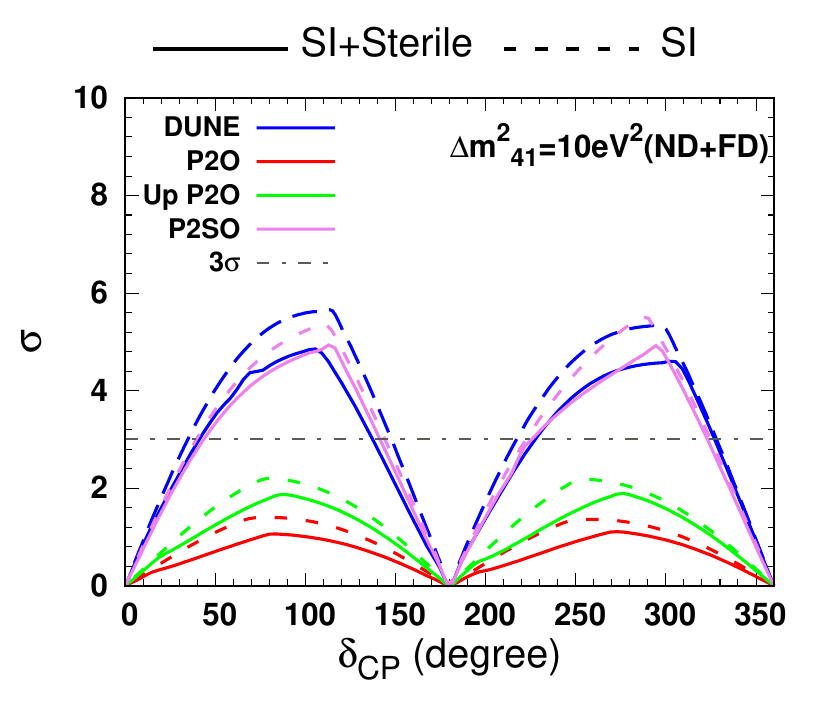} 
\includegraphics[width=0.49\textwidth]{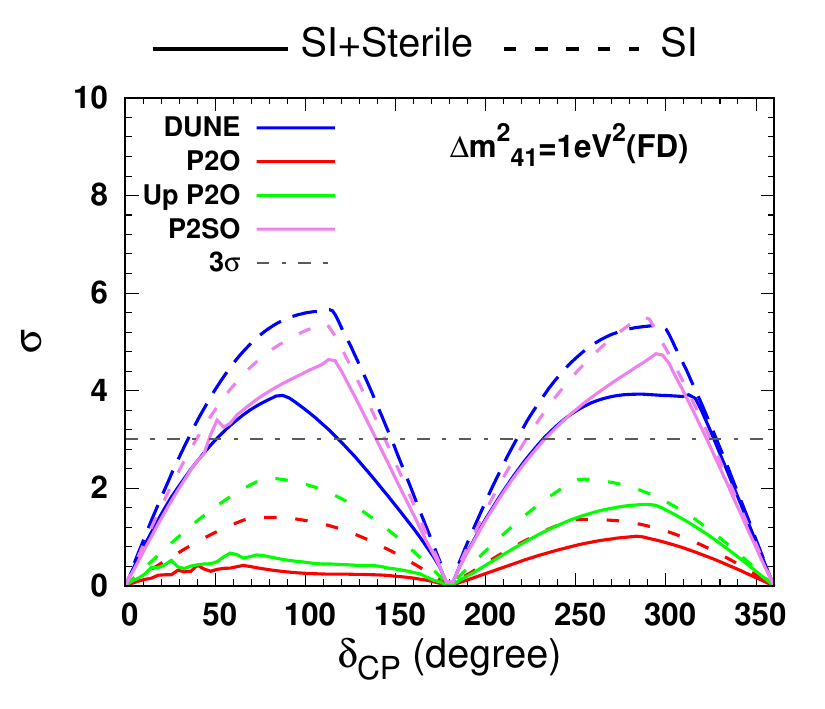}
\includegraphics[width=0.49\textwidth]{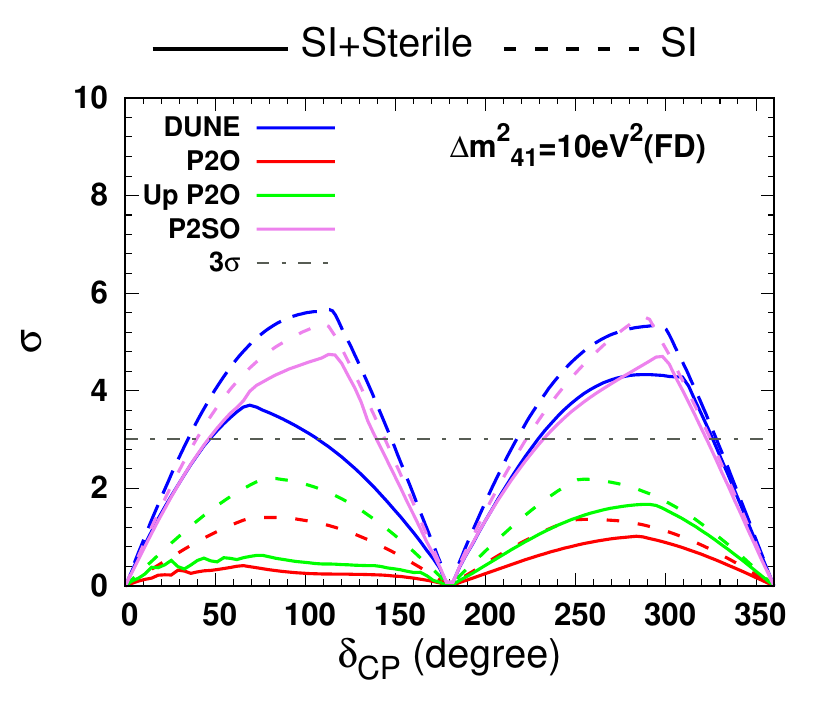}
\end{center}
\caption{CPV sensitivities as a function of true $\delta_{\rm CP}$ for DUNE, P2O, Upgraded P2O (Up P2O) and P2SO experiments in Standard Interaction (SI) case and SI + Sterile case for sterile neutrinos of masses $\Delta m^{\rm 2}_{4 1} =$  1 eV$^2$ and 10 eV$^2$.}
\label{fig_cpv}
\end{figure}

The oscillation parameter $\delta_{\rm CP}$ is bit slow as compared to other oscillation parameters in the race of precise measurement of them. The parameter $\delta_{\rm CP}$ has great importance to explain the CP violation in neutrino sector and also can explain the matter and antimatter asymmetry of the universe. In order to show the potential to find CP violation in nature, we have calculated CP violation (CPV) sensitivities for different long-baseline experiments. CPV sensitivity shows the ability of  the experiment to exclude the CP conserving values of $\delta_{\rm CP}$, if CP violation exists. In order to calculate CP violation sensitivity, we obtain $\chi^2$ assuming the CP conserving values in test and minimizing over them. Fig. \ref{fig_cpv} displays the CPV sensitivities as a function of true $\delta_{\rm CP}$ for standard and in presence of sterile neutrino in dashed and solid curves, respectively. Plots on the left panel of the figure show the sensitivities with sterile neutrino of $\Delta m^2_{41}=$ 1 eV$^2$ while the right panel is for sterile neutrino of $\Delta m^2_{41}=$ 10 eV$^2$. Upper panel (lower panel) represents the CPV sensitivities with ND+FD (FD). In each plot the different colored curves, i.e., blue, red, green and magenta represent the CPV sensitivities for DUNE, P2O, Upgraded P2O and P2SO experiments, respectively.

As expected, in SI scenario (dashed curve) higher CPV sensitivity values are obtained around the maximum CP violating values ($90^0$ and $270^0$) of $\delta_{\rm CP}$ and lowest sensitivities are at CP conserving values. The long-baseline experiments P2O and Upgraded P2O are less significant towards CP violation as the sensitivities are below $3\sigma$ for the whole region of $\delta_{\rm CP}$. The reason for lower CPV sensitivities is the low background rejection capability of both the experiments. DUNE and P2SO experiments are more sensitive towards CP violation and can able to establish CP violation at $3\sigma$ C.L., if it exists in nature. Addition of sterile neutrino to three neutrino scenario results to decrease in minimum CPV sensitivities (solid curves) for all the experiments. Also, effect of ND in each experiment is clearly visible. After including ND to FD, significant increase in sensitivities is observed, compared to only FD in each experiment in $3+1$ scenario for $\delta_{\rm CP}$ in the region $0^0$ to $180^0$. We can see that, in presence of sterile neutrino there is no significant improvement of CPV sensitivity in the region $180^0$ to $360^0$ of $\delta_{\rm CP}$ after adding ND to the FD. This feature can be explained in a similar fashion as the octant case. Note that the CP violation in presence of sterile neutrino is not only due to the standard CP phase, but  can also be due to the sterile phases. By varying the $\delta_{24}$ and $\delta_{34}$ in their allowed ranges, one can get bands instead of a single curve, by calculating maximum and minimum values of $\Delta \chi^2_{\rm min}$  as shown in Ref~\cite{Dutta:2016glq}. Sterile phases can enhance or deteriorate CPV sensitivity depending upon their values. In the next section, we have shown the effects of $\delta_{24}$ on the measurement of $\delta_{\rm CP}$. 

\section{Constraining CP phases}
\label{cpp}

\begin{figure}[hbt!]
\begin{center}
\includegraphics[width=0.47\textwidth]{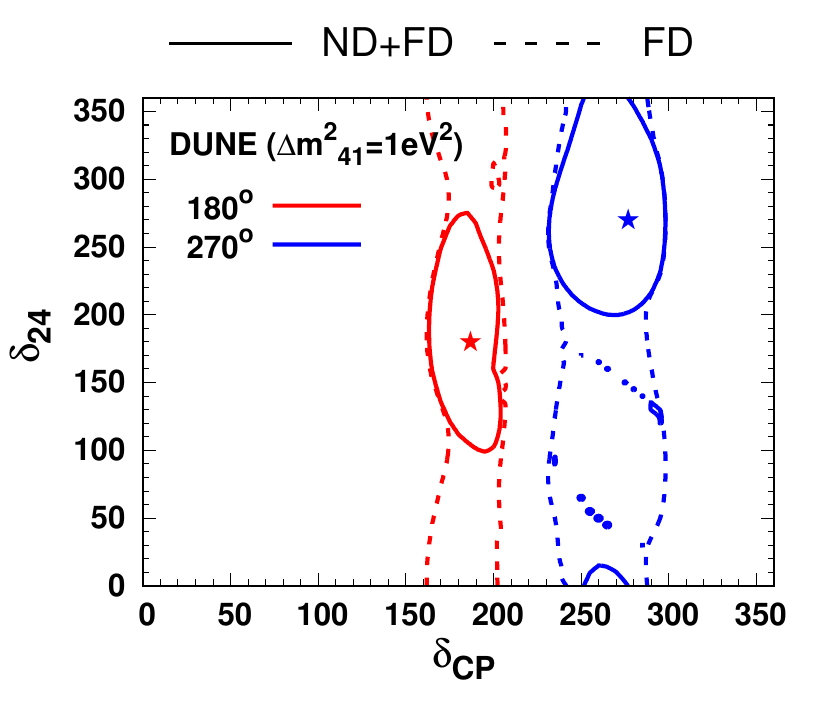}
\includegraphics[width=0.47\textwidth]{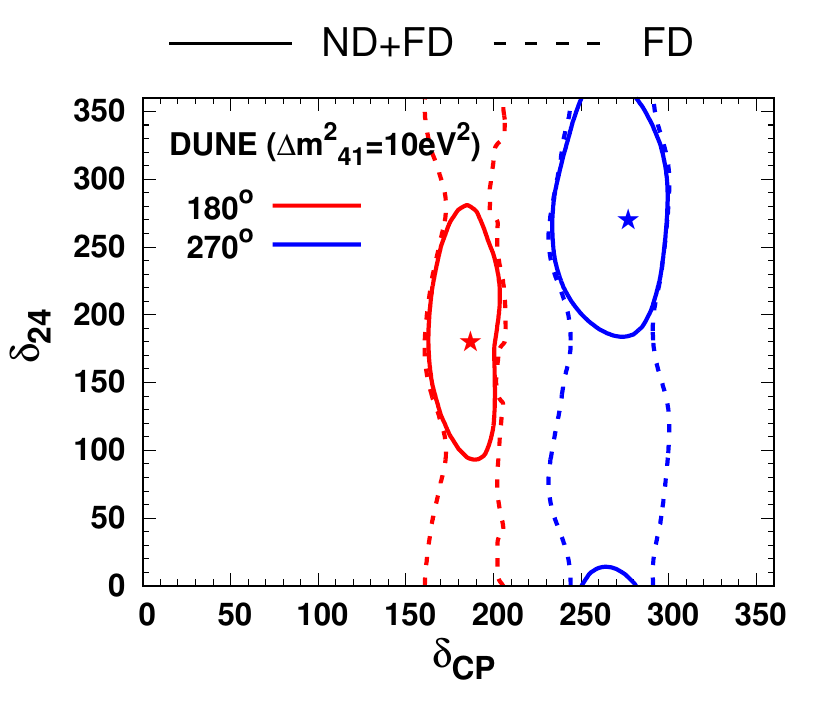} 
\includegraphics[width=0.47\textwidth]{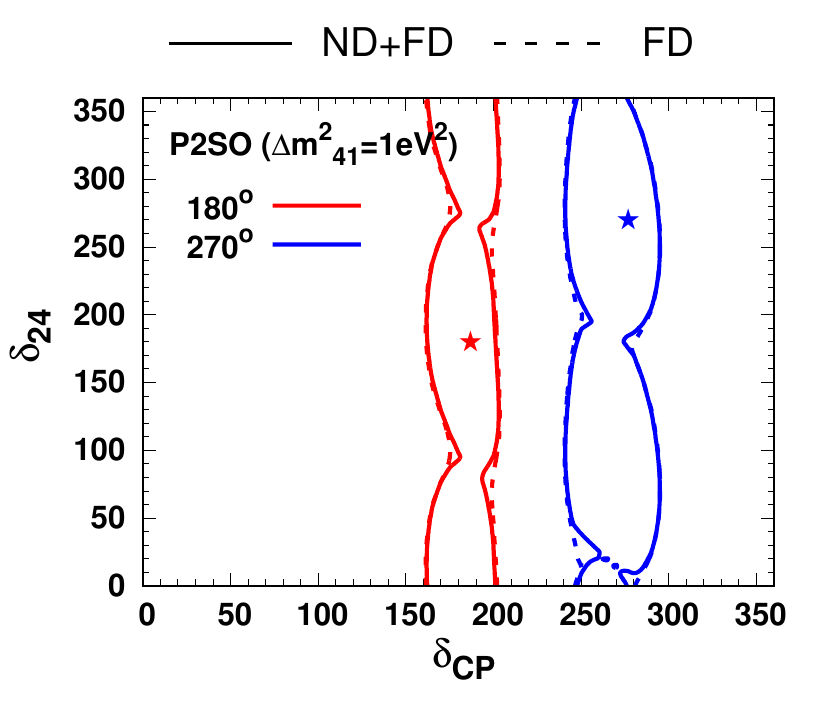}
\includegraphics[width=0.47\textwidth]{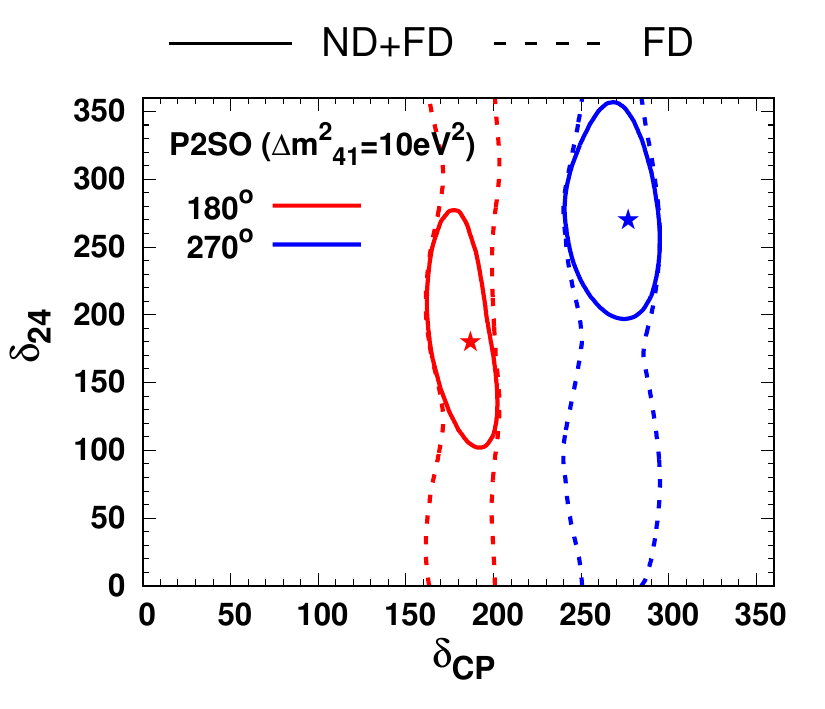}
\end{center}

\caption{CP precision sensitivity in $\delta_{\rm CP}$ and $\delta_{24}$ plane  for DUNE and P2SO experiments. We have shown the sensitivity at 1 $\sigma$ C.L. in presence of a light sterile neutrino of masses $\Delta m^{\rm 2}_{4 1} = 1$ eV$^2$ and $10$ eV$^2$.}
\label{fig_cpp}
\end{figure}

As discussed in the previous section, additional CP phases can be the reason for CP violation. In this section, we have shown CP precision sensitivities for long baseline experiments DUNE and P2SO as shown in Fig. \ref{fig_cpp}.  In our study, we have not considered P2O and Upgraded P2O because they have limited sensitivity to $\delta_{\rm CP}$. 
CP precision sensitivity says how precisely one can measure a particular value of CP phase. In our analysis, we have estimated the CP precision sensitivity, considering CP conserving or maximal CP violating values for the CP phases $\delta_{\rm CP}$ and $\delta_{24}$ i.e., for the true combination of  ($\delta_{\rm CP}$, $\delta_{24}$) of ($180^\circ$, $180^\circ$) and ($270^\circ$, $270^\circ$). Note that CP precision in 3+1 scenario can be studied for many combinations of ($\delta_{\rm CP}$, $\delta_{24}$). Our choice for the values of CP phases is motivated by the recent measurement of $\delta_{\rm CP}$ by T2K \cite{con_talk_t2k} and NO$\nu$A \cite{con_talk}. The measurement of T2K gives a best-fit value of $\delta_{\rm CP}$ around the maximal CP violating value of $270^\circ$ whereas the current best-fit value of this parameter by NO$\nu$A is around the CP conserving value of $180^\circ$. Marginalisation has been done over all standard and sterile oscillation parameters except the mass squared difference $\Delta m^2_{41}$. Left (right) panel of the figure shows the sensitivity with sterile neutrino  mass order of 1 eV$^2$ (10 eV$^2$). Upper panel is for DUNE experiment while lower panel is for P2SO experiment. In each plot, the dashed (solid) curve represents the allowed region between $\delta_{\rm CP}$ and $\delta_{\rm 24}$ at $1\sigma$ C.L. considering FD (ND+FD). Red and blue colour of the curves represent the true value of CP phases $\delta_{\rm CP}$ and $\delta_{24}$ as $180^\circ$ and  $270^\circ$, respectively. Star marks in each plot show the true point considered for analysis. 

As expected the inclusion of ND to FD  increases the sensitivities of the experiments as compared to only FD except for the case of P2SO and $\Delta m^2_{41} = 1$ eV$^2$. Due to the effect of ND, parameter spaces  (specially the sterile angles) are more constrained and experiments can measure the CP conserving and CP violating values of the CP phases with great accuracy. For P2SO and $\Delta m^2_{41} = 1$ eV$^2$, addition of ND does not improve the sensitivity.  This is because, in this case, the flux is optimal for $\Delta m^2_{41} = 10$ eV$^2$ at the ND. For $\Delta m^2_{41} = 10$ eV$^2$, both the experiments can equally measure the CP violating and CP conserving values of the phases. Note that for the measurement of phases in 3+1 scenario, it would be ideal to have a ND  capable of detecting $\nu_\tau$ via $\nu_\mu \rightarrow \nu_\tau$ \cite{Donini:2008wz}. However, even without the capability of $\tau$ neutrino detection, having a replica of the far detector at a short distance, one can keep the systematics under control and measure the phases at a good precision.


\section{Summary and conclusions}
\label{sum}

In this paper we have studied the capability of the long-baseline experiment options at the KM3NeT facility to probe the light sterile neutrino. In our study, we have considered  the options P2O, which will have neutrinos from a 90 KW beam to be detected at the ORCA detector, the Upgraded P2O, which will have neutrinos from the upgraded 450 KW beam to be detected at the ORCA detector and the option of P2SO, which will have neutrinos from a 450 KW beam to be detected at the upgraded Super-ORCA detector. All these options will have a baseline around 2595 km. We have also compared the results of these three options with DUNE. 

In our study, we showed that the options at the KM3NeT is more sensitive if the value of $\Delta m^2_{41}$ is around 10 eV$^2$ whereas the DUNE is more sensitive if the value of $\Delta m^2_{41}$ is around 1 eV$^2$. Our results also show that near detector is very important for the study of sterile neutrinos and addition of near detector improves the sensitivity as compared to only far detector for 3+1 scenario. Among the three options at KM3NeT, the sensitivities of P2O and upgraded P2O are limited because of their poor background rejection capabilities. Regarding the capability of constraining the sterile mixing parameters, we find that both P2SO and DUNE have good efficiency to constrain the mixing angles $\theta_{14}$ and $\theta_{24}$. DUNE is better for $\Delta m^2_{41} = 1$ eV$^2$ while P2SO is better for $\Delta m^2_{41} = 10$ eV$^2$. The sensitivities of P2SO and upgraded P2O are exactly same for ND+FD configuration. Regarding determination of the unknowns in 3+1 scenario, we find that the sensitivities in presence of sterile neutrino is lower as compared to standard three flavour case. Addition of ND improves the sensitivities but still they are less than the standard three flavour case. In general, we have observed that, for $\Delta m^2_{41} = 10$ eV$^2$, improvement in the sensitivity due to the addition of ND is more for the long-baseline experiment options at KM3NeT as compared to $\Delta m^2_{41} = 1$ eV$^2$. Regarding hierarchy sensitivity in presence of sterile neutrinos, sensitivity of Upgraded P2O is similar to DUNE whereas P2SO gives the best sensitivity for ND+FD. Regarding octant and CP sensitivities, we find that the sensitivities of P2O and Upgraded P2O are poor compared to DUNE and P2SO in the 3+1 scenario and for ND+FD. For octant, the best sensitivity comes from P2SO and  for $\delta_{\rm CP}$, DUNE provides slightly better sensitivity than P2SO regarding both CP violation and CP precision. We find that addition of ND does not play any role in the case of (i) octant sensitivity in the higher octant for P2SO, (ii) CPV sensitivity in the $180^0$ to $360^0$ region of $\delta_{CP}$ for all the experimental setups and  (iii) CP precision sensitivity for $\Delta m^2_{41} = 1 $ eV$^2$ for P2SO. 

In summary, our results show that the sensitivity of P2SO, i.e., the long-baseline option of KM3NeT where neutrinos from a 450 KW beam at Protvino will be detected at the Super-ORCA detector, is either comparable or better than DUNE in both standard and 3+1 scenario. Our results will help in the design of the long-baseline experiments at KM3NeT facility.

\section*{Acknowledgements}

DKS acknowledges Prime Minister's Research Fellowship, Govt. of India. MG acknowledges Ramanujan Fellowship of SERB, Govt. of India, through grant no: RJF/2020/000082. RM acknowledges the support from University of Hyderabad through the IoE project grant IoE/RC1/RC1-20-012. We acknowledge the use of CMSD HPC facility of Univ. of Hyderabad to carry out computations in this work. This work has been in part funded by Ministry of Science and Education of Republic of Croatia grant No. KK.01.1.1.01.0001.

\bibliographystyle{JHEP}
\bibliography{sterile}
  
\end{document}